\documentclass[12pt]{iopart}
\pdfoutput=1

\usepackage{graphicx}
\usepackage{subfigure}
\usepackage{latexsym}   
\usepackage{bm}
\usepackage{amstext}

\usepackage[toc,page]{appendix}

\usepackage{etoolbox}
\usepackage[colorlinks=true, linkcolor=blue, citecolor=blue]{hyperref}

\usepackage{filecontents}

\usepackage{lipsum}

\makeatletter
\def\@mkboth#1#2{}
\newlength\appendixwidth
\preto\appendix{\addtocontents{toc}{\protect\patchl@section}}
\newcommand{\patchl@section}{%
  \settowidth{\appendixwidth}{{\bf \sffamily Appendix }}%
  \addtolength{\appendixwidth}{1.5em}%
  \patchcmd{\l@section}{1.5em}{\appendixwidth}{}{\ddt}%
}
\makeatother

\renewcommand{\abstractname}{{\bf \sffamily Abstract}}

\renewcommand{\figurename}{{\bf \sffamily Figure}}

\renewcommand{\tablename}{{\bf \sffamily Table}}

\begin{document}

\title[Kinetics of first-order phase transitions from microcanonical thermostatistics]{{\sffamily \huge Kinetics of first-order phase transitions from microcanonical thermostatistics}}

\author{\sffamily \large L. G. Rizzi$^{1}$}

\address{\sffamily $^1$Departamento de F\'isica, Universidade Federal de Vi\c{c}osa, Av.\,P.\,H.\,Rolfs,~s/n,~36570-900,~Vi\c{c}osa-MG,~Brazil.}
\ead{lerizzi@ufv.br}
\vspace{10pt}
\begin{indented}
\item[]\today
\end{indented}

\begin{abstract}
	More than a century has passed since van't Hoff and Arrhenius formulated their celebrated rate theories, but there are still elusive aspects in the temperature-dependent
phase transition kinetics of molecular systems.~Here
	 I present a theory based on microcanonical thermostatistics that establishes a simple and direct temperature dependence for all rate constants, including the forward and the reverse rate constants, the equilibrium constant, and the nucleation rate.~By
	considering a generic model that mimic the microcanonical temperature of molecular systems in a region close to a first-order phase transition, 
I obtain shape-free relations between kinetics and thermodynamics physical quantities 
which are validated through stochastic simulations.~Additionally, 
	the rate theory is applied to results obtained from protein folding and ice nucleation experiments, demonstrating that the expressions derived here can be used to describe the experimental data of a wide range of molecular systems.

%
\end{abstract}


%
\vspace{2pc}

%
%
%
%

\maketitle

\newpage

\title[Kinetics of first-order phase transitions from microcanonical thermostatistics]{~}

\noindent
\hrulefill

\tableofcontents

\vspace{0.5cm}

\noindent
\hrulefill


\section{Introduction}

	The temperature dependence of first-order phase transitions kinetics in molecular
systems is probably one of the oldest unsolved issues in modern science,
despite its importance to a wide range of processes in 
biology, climate, and materials science.~Examples
include the aggregation of misfolded or intrinsically disordered proteins, which is a phenomenon
that can be related to a number of proteinopathies~\cite{knowles2014natrev}~(e.g., Alzheimer's disease and type 2 diabetes); 
and the formation of protein crystals, which are 
used by crystallographic methods in the characterization of their
tridimensional structures~\cite{vekilov2016}.

	Misunderstandings may have arisen partially due to many Arrhenius-like 
expressions presented in the literature
which are often used indiscriminately beyond their scope~\cite{zhou2010review}.
	It is common to find studies 
where the rates are defined by expressions proportional to $e^{-\Delta G^{\ddagger}/k_BT}$ with effective activation energies $\Delta G^{\ddagger}$ that are independent of the temperature $T$, just as those derived by Eyring~\cite{eyring1935jcp} and Kramers~\cite{kramers1940phys} which underpin the well-known transition state theories~\cite{zhou2010review}. 
	The problem is that, if $\Delta G^{\ddagger}$ does not display any dependence on $T$, the corresponding rate will increase with temperature, 
however this is exactly the opposite behaviour observed in processes which take place in many  finite molecular systems, e.g., folding of heteropolymers and molecular crystallization.
	Remarkably, ``anomalous''  behaviours like that are known for decades, but it seems that they are still causing confusion today~\cite{cooper2010jphyschemlett}.

	Of particular interest to first-order phase transitions is 
the temperature-dependent kinetics of nucleation processes, 
which is characterized by the nucleation rate $j(T)$.
	Usually, the nucleation rate is defined by an Arrhenius-like expression as
\begin{equation}
j(T) = A \, e^{-\Delta G^*/k_B T} ~~,
\label{CNT_nucleation_rate}
\end{equation}
where $\Delta G^*$  can be thought as an energetic barrier that is needed to be overcomed
through fluctuations in order to convert part of the system to the new phase 
and is related to work to form the critical nucleus~\cite{dimobook,schmelzerbook};
and $A$ is a pre-factor that determines the dimensional units of the nucleation
rate and is related to the attachment frequencies to the critical nucleus and to 
the Zeldovich factor~\cite{dimobook}.~Equation~(\ref{CNT_nucleation_rate})
	was first introduced by Volmer \& Weber and, 
together with the seminal contributions made by Becker \& D\"oring, Farkas, Frenkel, 
Gibbs, Kaischew, Stranski, Tammann, and Zeldovich,
it forms the basis of what we know as the classical nucleation theory (CNT)~\cite{dimobook}.
	Unlike the aforementioned Arrhenius-like expressions related to the rate constants,
there were several attempts to include the temperature dependence on $A$ and $\Delta G^{*}$ defined in equation~(\ref{CNT_nucleation_rate}).
	Although successful approaches indicate that $\Delta G^{*}$ should increase as $T$ approaches the transition temperature~\cite{dimobook,schmelzerbook},
quantitative agreement to experimental data is rarely observed.




	A fundamental drawback of nucleation theories that are based on CNT
and which attempt to describe first-order phase transition kinetics in molecular systems
is their need for assumptions based on geometric features of the system, e.g.,~that the critical nucleus should have a spherical shape.~References~\cite{cabriolu2012jcp,bingham2013jcp} 
	show that this type of assumption is particularly problematic, specially for systems with anisotropically interacting molecules.

	Interestingly, many years ago, Schmelzer {\it et al}.~\cite{schmelzerbook}
have suggested that an approach based on microcanonical thermostatistics~\cite{grossbook} could have provided an alternative nucleation theory.
	Nevertheless, only recently that idea has gained attention with the works of Janke and collaborators.~For 
	instance, Zierenberg {\it et al.}~\cite{janke2017natcomm} explored
the shape-free properties of  microcanonical free-energy profiles to describe the aggregation of polymeric chains without making any assumptions about the geometry of the critical nuclei.~A few years before, a similar approach based on microcanonical thermostatistics was used
to infer the kinetics of polymeric chains close to their folding-unfolding transitions~\cite{frigori2013jcp}.
	As it happens, the authors of references~\cite{janke2017natcomm,frigori2013jcp} 
(including myself) missed the opportunity to identify any temperature dependence on the effective 
activation energies that were evaluated from the microcanonical free-energy profiles.

	In this paper I recall that idea in order to develop a rate theory that is based on the microcanonical thermostatistics~\cite{grossbook}.
	By doing so, I am able to establish a straightforward relationship between 
the thermodynamics and the kinetics of first-order phase transitions in molecular systems.
	In particular, I derive  simple temperature-dependent expressions for all rates, including the nucleation rate, which are validated by both numerical simulations and
experimental data on protein folding and ice nucleation phenomena.





\section{Microcanonical thermostatistics}
\label{micro_thermostat}

	First, I clarify that the rate theory introduced here
is different from both microcanonical transition state~\cite{truhlar2001pnas}
and adiabatic nucleation~\cite{meyer1986jcrysgrow} 
theories.~Also, 
	I emphasize that, although the textbook meaning of
{\it microcanonical} is that the system have a constant energy and is completely isolated, 
here it is used mainly because the approach 
is based on the density of states, which is a quantity that
describes fundamental properties of the system
regardless its coupling to a thermal reservoir.~In fact, 
	as my aim is to study temperature-driven
first-order phase transitions, the terminology microcanonical 
thermostatistics is used only to indicate that I am interested in
intrinsically finite systems~\cite{grossbook,schnabel2011pre}, 
e.g.,~molecular systems either with aggregating molecules or with folding heteropolymers.
	In that case, one should consider that the molecular system have a fixed number 
of molecules $N$ 
and a constant volume $V$, so that its microcanonical entropy is given by
\begin{equation}
S(E) = k_B \ln\Omega(E) + C ~~,
\end{equation}
where $\Omega(E)$ is the density of states, i.e.,~the number of microscopic
configurations which have energy between $E$ and $E+\varepsilon$; 
$k_B$ is the Boltzmann's constant;
and $C$ is an arbitrary constant which depends on the bin size $\varepsilon$ and 
on a reference value for the entropy.~Importantly, 
	the microcanonical analysis is usually restricted to cases where $E$ denotes the
internal energy of the system, i.e., the sum of potential and kinetic energies of the
molecules in the system (eventually excluding the energy of explicit solvent molecules or particles that are in a reservoir), 
just as those defined in molecular dynamics simulations~\cite{frenkelbook}.~One 
	should note that, because $N$ and $V$ are fixed, changes in the internal 
energy $E$ and enthalpy $H$ would be equivalent, and variations in the
Helmholtz free energy $F$ would be equal to changes in the Gibbs potential $G$.
	Also, just for the moment, I assume that $k_B=1$ in order to display the numerical
 results in units comparable to those usually obtained from simulations
which explore the microcanonical thermostatistics as an analysis method.
	It is worth mentioning that the microcanonical entropy 
$S(E)$ can be conveniently obtained by several computational algorithms, e.g., 
multicanonical~\cite{berg2003cpc},
broad histogram~\cite{pmc1996brazj},
Wang-Landau~\cite{wanglandau2001prl}, and
statistical temperature weighted histogram method~\cite{straub2011jcp,rizzi2011jcp}.

\begin{figure}[!b]
\centering
\includegraphics[width=0.52\textwidth]{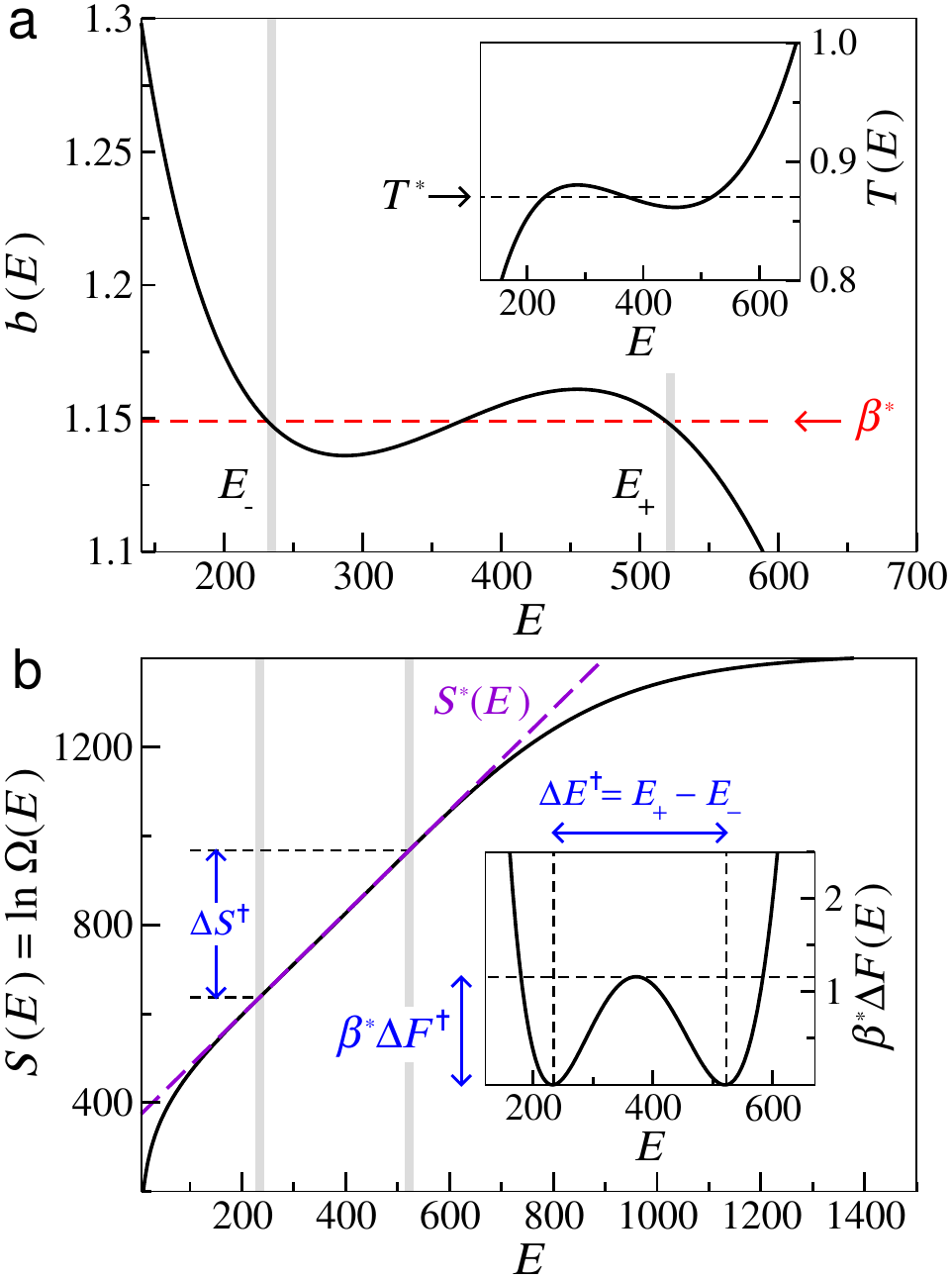}
\caption{
{\bf Microcanonical thermostatistics.}
(a) Displays the inverse of the microcanonical temperature $b(E)$ which is characteristic of molecular systems that present first-order phase transitions (see~\ref{micromodel} for details).
Dashed (red) line indicate a Maxwell-like construction used to determine the inverse of the transition temperature 
$\beta^{*}=1.1489$.  
Inset: microcanonical temperature $T(E)=1/b(E)$ obtained by equation~(\ref{tempE}), which
present a transition temperature equal to
$T^{*}=1/\beta^{*}=0.8704$ (here it is assumed that $k_B=1$).
(b) Continuous (black) and dashed (violet) lines correspond to the microcanonical entropy $S(E)$ and 
its linear approximation (at the transition), $S^{*}(E) = \beta^{*} E - \beta^{*} F(E_{-})$, with $F(E_{-})=E_{-} - T^{*}S(E_{-})$, respectively.
Inset: microcanonical free-energy profile $\beta^{*} \Delta F(E)$ with minima at energies
$E_{-}=232$ and $E_{+}=520$, which are indicated by the vertical (grey) lines.
Blue arrows indicate the free-energy barrier 
$\beta^{*} \Delta F^{\dagger}=1.16$ at the energy $E^{*}=372$,
the microcanonical latent heat, $\Delta E^{\dagger}=288$,
and the microcanonical entropy difference, $\Delta S^{\dagger}=331$.
}
\label{microcanonical_thermostatistics}
\end{figure}

	One important intensive quantity that can be obtained from the microcanonical 
entropy is the inverse of the microcanonical temperature, which is given by
\begin{equation}
b(E)=\frac{dS(E)}{dE} ~~ ~ .
\label{betaE}
\end{equation}
	Just like the density of states $\Omega(E)$, both $b(E)$ and the microcanonical
temperature $T(E)=1/b(E)$ can be used to describe fundamental properties of the system and 
are uniquely determined by $N$, $V$, and $E$.~Regardless
	if the system display, e.g., homogeneous or heterogeneous nucleation, 
or polymeric folding and/or aggregation phenomena, first-order phase transitions are 
generically characterized by the presence of a S-shaped curve
in the inverse of the microcanonical temperature~\cite{grossbook,schnabel2011pre}, 
as showed in figure~\ref{microcanonical_thermostatistics}(a).
	In order to explore the generic aspects of my approach
I evaluate both $b(E)$ and $S(E)$ from a model function that mimic 
the microcanonical temperature $T(E)$ of molecular systems
near a first-order phase transition (see~\ref{micromodel}).

	As shown in figure~\ref{microcanonical_thermostatistics}(a), 
the transition temperature $T^*=1/\beta^*$ can be estimated from
$b(E)$ through a Maxwell-like construction~\cite{janke2017natcomm,schnabel2011pre}.
	Equivalently, the inverse of the transition temperature
can be determined as
\begin{equation}
\beta^{*}= \frac{\Delta S^{\dagger}}{\Delta E^{\dagger}} ~~~ ,
\label{transition_temperature_dagger}
\end{equation}
where $\Delta E^{\dagger}=E_{+}-E_{-}$ denotes the energy difference, or the microcanonical latent heat, between the phases at energies $E_{+}$ and $E_{-}$, while $\Delta S^{\dagger} =S(E_{+}) - S(E_{-})$ is the microcanonical entropy difference,
as illustrated in figure~\ref{microcanonical_thermostatistics}(b).
	At the transition, one can consider
$\beta^{*} F(E) = \beta^{*} E - S(E)$, so that the microcanonical free-energy 
profile is defined as
\begin{equation}
\beta^{*} \Delta F(E) = \beta^{*} [ F(E) - F(E_{-})] = S^{*}(E) - S(E)~~~,
\label{free_energy_profileEq}
\end{equation}
with $S^{*}(E) = \beta^{*}(E-E_{-})+S(E_{-})$.
	Although the linear function $S^{*}(E)$ looks very similar to the microcanonical 
entropy in figure~\ref{microcanonical_thermostatistics}(b), there is a convex intruder region in $S(E)$ between the energies $E_{-}$
and $E_{+}$, which is evidenced by the free-energy profile showed in the inset
of figure~\ref{microcanonical_thermostatistics}(b).
	Importantly, the microcanonical free-energy profile given by 
equation~(\ref{free_energy_profileEq}) can be directly related to the equal height canonical
probability distribution at $\beta^{*}$~\cite{janke1998npB,leekosterlitz1990prl}, i.e.,
$p^{*}(E) \propto e^{-\beta^{*} \Delta F(E)}$.




\section{First-order phase transition kinetics}

	In order to study the kinetics of the system, 
I devised a stochastic protocol in~\ref{stochasticsim} that 
allows one to simulate the interaction between the system and a canonical
thermal reservoir at a constant temperature $T$,
 so that the stationary distribution
is equal to the canonical equilibrium distribution at $\beta=1/k_BT$ (see figure~\ref{simulations_series}), that is,
\begin{equation}
p(E) = \frac{\Omega(E) \, e^{-\beta E}}{\mathcal{Z}(\beta)} ~~,
\label{canonical_prob}
\end{equation}
where one can evaluate the canonical partition function $\mathcal{Z}(\beta) = \sum_E \Omega(E) \, e^{-\beta E}$ by assuming that the density of states is given by $\Omega(E)=e^{S(E)}$, with $S(E)$ obtained from the inverse of the microcanonical temperature $b(E)$ as discussed in~\ref{micromodel}.
	In order to perform the simulations, 
I consider that the energy exchange between the system and the thermal reservoir
occur at random, just as in the stochastic processes defined in reference~\cite{crooks2000pre}, 
which are known to be applicable even for 
non-equilibrium processes 
(see~\ref{stochasticsim} for details).
	Also, because the rate constants are evaluated
directly from the mean first-passage times, the simulations are 
constructed in a way that the diffusion in the energy space allows 
one to reconstruct the time variable just as discussed in 
reference~\cite{krivov2013pre}.

\subsection{Rate constants and the equilibrium rate}
\label{rate_const_sec}

\begin{figure*}[!b]
\centering
\includegraphics[width=0.97\textwidth]{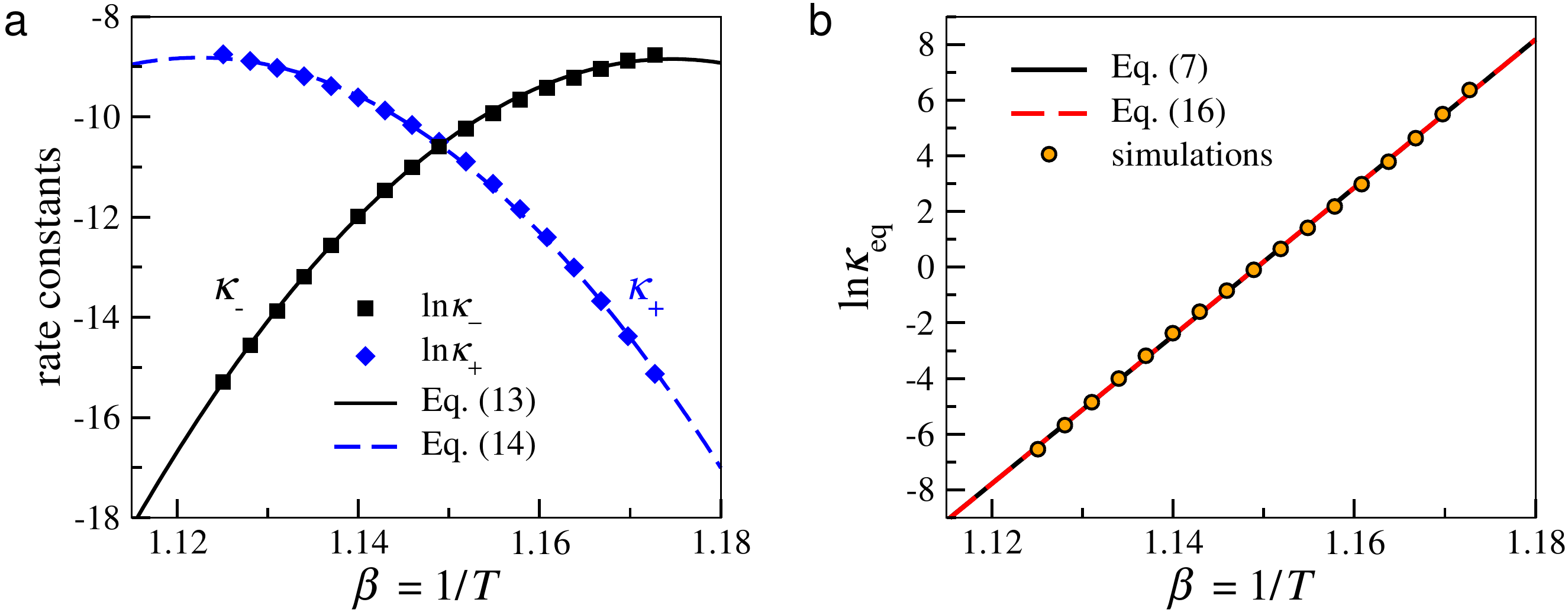}
\caption{
\footnotesize {\bf Rate constants and the equilibrium rate.} 
(a) Arrhenius plots for the forward $\kappa_{-}$ and the reverse $\kappa_{+}$ rate constants.
	Filled symbols correspond to numerical results obtained from the stochastic simulations
and lines denote the fit to equations~(\ref{kappa_minus}) and~(\ref{kappa_plus})
considering $\beta^{*} \Delta F^{\dagger}=1.16$ and $\beta^{*}=1.1489$.
	The parameters obtained for $\kappa_{-}$ are
$A_{+}=2.54\times 10^{-5}$, $\Delta E_{+}^{\ddagger}=-135.2$, and $\bar{\gamma}_{+}=5262.6$;
while for $\kappa_{+}$ are
$A_{-}=2.62\times 10^{-5}$, $\Delta E_{-}^{\ddagger}=130.8$, and $\bar{\gamma}_{-}=4940.9$.
	By considering $\tau_{\varepsilon}=1$ and $\varepsilon=2$,
both pre-factors $A_-$ and $A_+$ defined by
equation~(\ref{prefactors}) yield $\gamma^{*}\approx 2 \times 10^{-4}$. 
	(b) Arrhenius plot for the equilibrium constant $\kappa_{\text{eq}}=\kappa_{-}/k_{+}$. 
	Circles correspond to results 
that were computed from the numerical data
showed in (a).
	Straight (black) line denote the linear regression using equation~(\ref{kappa_eq_vanthoff}), from where one have $\Delta E^{\ominus} = 266$ and $\Delta S^{\ominus} = 305.6$; while the dashed (red) line correspond to equation~(\ref{kappa_eq})
by assuming that $\Delta E^{\ddagger} = \Delta E_{+} - \Delta E_{+}^{\ddagger}=266$
and $\Delta \bar{\gamma} = \bar{\gamma}_{-} - \bar{\gamma}_{+} = -321.7$.
	The stochastic simulations were performed as described in~\ref{stochasticsim} 
and considering the microcanonical entropy $S(E)$ showed in figure~\ref{microcanonical_thermostatistics}
(where it is assumed that $k_B=1$).
}
\label{fig_rate_constants}
\end{figure*}

	Figure~\ref{fig_rate_constants}(a) shows Arrhenius plots for the forward ($\kappa_{-}$) and the reverse ($\kappa_{+}$) rate constants, which are obtained respectively from the inverse of the mean first-passage times~\cite{hanggi1999pre}, $\tau_{-}$ and $\tau_{+}$, that were  numerically evaluated from the stochastic simulations through the method of labelled walkers mentioned in reference~\cite{trebst2004pre}
(see figure~\ref{simulations_series} in~\ref{stochasticsim} for details).
	Accordingly, as the temperature $T=1/\beta$ decreases, the forward rate constant $\kappa_{-} = 1/\tau_{-}$ increases, which means that the mean first-passage time $\tau_{-}$ for the system to go from $E_{+}$ to $E_{-}$ decreases
with the temperature, so that more of those events will occur per unity of time.
	The result in figure~\ref{fig_rate_constants}(b) show that, although $\kappa_{-}$ (and $\kappa_{+}$) display a behaviour that is typical of non-Arrhenius kinetics, the logarithmic of the equilibrium constant, $\kappa_{\text{eq}} = \kappa_{-} / \kappa_{+}$,
might display a remarkably linear behaviour as function of $\beta$.
	By considering the fit to the numerical data of the expression obtained from the van't Hoff equation, that is,
\begin{equation}
\kappa_{\text{eq}}^{\ominus}
=
\exp
\left[
\beta \Delta E^{\ominus} - \Delta S^{\ominus}
\right]~~,
\label{kappa_eq_vanthoff}
\end{equation}
one finds
 $\Delta E^{\ominus}=266$ and $\Delta S^{\ominus}=306$,
so that $T^{*} = (\beta^{*})^{-1} \approx (\Delta S^{\ominus}/\Delta E^{\ominus})^{-1} = 0.87$. 
	Interestingly, as suggested by the analytical expressions derived in the following, 
the enthalpic ($\Delta E^{\ominus}$) and the entropic ($\Delta S^{\ominus}$)
contributions obtained from equation~(\ref{kappa_eq_vanthoff})
should be related, respectively, to the values of $\Delta E^{\dagger}$ and 
$\Delta S^{\dagger}$ evaluated from the microcanonical 
thermostatistics analysis presented in figure~\ref{microcanonical_thermostatistics}(b).

	As described in~\ref{MFPTapp}, a straightforward way to obtain temperature-dependent expressions for the rate constants
from the microcanonical entropy $S(E)$ is  to evaluate the 
mean first-passage times from the canonical equilibrium distribution given by equation~(\ref{canonical_prob}).
	Hence, as I discuss below, both $\tau_{-}$ and $\tau_{+}$ can be obtained
from the estimates for
$p(E)$ at energies close to $E_{-}$, $E_{+}$, and $E^{*}$, which are obtained from approximated expressions for the density of states $\Omega(E)$.
	Indeed, by considering the free-energy profile 
defined by equation~(\ref{free_energy_profileEq}), one can
estimate $\Omega(E)$ directly from the microcanonical entropy which is given by
 $S(E)= S^*(E)-\beta^{*}\Delta F(E)$.
	In particular, the expansion of the free-energy profile around its
maximum at the energy $E^{*}$ can be written as
\begin{equation}
\beta^* \Delta F(E) \approx \beta^* \Delta F^{\dagger} -
 \frac{\gamma^{*}}{2}(E-E^{*})^2~~,
\label{betadeltaF_ast}
\end{equation}
where $\gamma^{*} \approx \left. (db(E)/dE)\right|_{E=E^{*}}$ is a positive constant.
	Thus, the canonical probability distribution around the energy $E^{*}$ and at a inverse temperature $\beta$ close to $\beta^{*}$ can be estimated as
\begin{equation}
p^{*}(E) \approx \frac{\Gamma^{*}(\beta)}{\mathcal{Z}(\beta)}
\exp
\left\{ \frac{\gamma^{*}}{2} \left[ (E - E^{*}) - \frac{(\beta - \beta^{*})}{\gamma^{*}} \right]^2 \right\}
~~,
\label{prob_ast}
\end{equation}
with 
\begin{eqnarray}
\hspace{-1.0cm}
\Gamma^{*}(\beta) = 
\exp
\biggl[
S(E_{\pm}) - \beta^{*}E_{\pm}
- \beta^{*} \Delta F^{\dagger} 
\biggr.
 \left.
- \frac{((\beta - \beta^{*}) + \gamma^{*} E^{*} )^2}{2 \gamma^{*}}
+\frac{\gamma^{*} (E^{*} )^2}{2}
\right]
~ ,
\label{gamma_ast}
\end{eqnarray}
where the subscript in $E_{\pm}$ indicate that either $E_{-}$ or
$E_{+}$ can be used (as long as mixed notations are avoided).
	Similarly, by assuming that
$\beta^* \Delta F(E) \approx (\gamma_{\pm}/2)(E-E_{\pm})^2$,
with $b(E_{\pm}) \approx b(E^{*}) \approx \beta^*$ and 
$\gamma_{\pm} \approx \left. (db(E)/dE) \right|_{E=E_{\pm}}>0$,
the probability distribution close to the energy
$E_{-}$ (or $E_{+}$), and at $\beta$ close to $\beta^{*}$,
can be approximated by
\begin{equation}
p_{\pm}(E) \approx 
\frac{\Gamma_{\pm}(\beta)}{\mathcal{Z}(\beta)}
\exp\left\{
-\frac{\gamma_{\pm}}{2}
\left[
(E - E_{\pm}) + \frac{(\beta - \beta^{*})  }{\gamma_{\pm}}
\right]^2
\right\}
~,
\label{prob_plus_minus}
\end{equation}
with
\begin{eqnarray}
\Gamma_{\pm}(\beta) = 
\exp
\biggl[
S(E_{\pm}) - \beta^{*}E_{\pm} 
\biggr.
\left.
+ \frac{((\beta - \beta^{*}) - \gamma_{\pm} E_{\pm} )^2}{2 \gamma_{\pm}}
-\frac{\gamma_{\pm} (E_{\pm} )^2}{2}
\right]
~.
\label{gamma_plus_minus}
\end{eqnarray}
As shown in~\ref{MFPTapp}, 
the forward 
and the reverse rate 
constants can be obtained from 
equations~(\ref{prob_ast}),~(\ref{gamma_ast}),~(\ref{prob_plus_minus}), and~(\ref{gamma_plus_minus}), and 
are given, respectively, as
\begin{equation}
\kappa_{-} = \frac{1}{\tau_{-}}
\approx 
A_{+} \exp
\left[
- \Delta E_{+}^{\ddagger} (\beta - \beta^{*})
- \frac{\bar{\gamma}_{+}}{2} (\beta - \beta^{*})^2 
\right]~~,
\label{kappa_minus}
\end{equation}
\begin{equation}
\kappa_{+} = \frac{1}{\tau_{+}}
\approx 
A_{-} \exp
\left[
- \Delta E_{-}^{\ddagger} (\beta - \beta^{*})
- \frac{\bar{\gamma}_{-}}{2} (\beta - \beta^{*})^2 
\right]~~,
\label{kappa_plus}
\end{equation}
with
\begin{equation}
A_{\pm} =
\sqrt{ \frac{\varepsilon^4\gamma^{*}}{8 \pi \tau_{\varepsilon}^2 ( \Delta E_{\pm}^{\ddagger})^2}}
~e^{- \beta^{*} \Delta F^{\dagger}}~~~, 
\label{prefactors}
\end{equation}
where $\bar{\gamma}_{\pm} =  (\gamma_{\pm})^{-1} + (\gamma^{*})^{-1}$,
$\Delta E_{\pm}^{\ddagger} =  E^{*}-E_{\pm}$, and $\tau_{\varepsilon}$
is a characteristic time scale involved in a microscopic energy transition.~Also, 
	from equations~(\ref{kappa_minus}) and~(\ref{kappa_plus}), one can readily 
evaluate the equilibrium rate constant, which is given by
\begin{equation}
\kappa_{\text{eq}} = \frac{\kappa_{-}}{\kappa_{+}}
\approx 
A \,
 \exp
\left[
\Delta E^{\ddagger} (\beta - \beta^{*})
+ \frac{\Delta \bar{\gamma}}{2} (\beta - \beta^{*})^2 
\right]~~,
\label{kappa_eq}
\end{equation}
where 
$\Delta E^{\ddagger} = \Delta E_{-}^{\ddagger} - \Delta E_{+}^{\ddagger} =
E_{+} - E_{-}$, 
$A = \Delta E_{-}^{\ddagger} / |\Delta E_{+}^{\ddagger}|$, 
and 
$\Delta \bar{\gamma} = (\gamma_{-})^{-1} - (\gamma_{+})^{-1}$.

	As shown in figure~\ref{fig_rate_constants}(b), equation~(\ref{kappa_eq}) fits the numerical results obtained from the stochastic simulations very well and, by assuming that
$\beta^{*}=1.1489$, one finds that 
$\Delta \bar{\gamma} = -321.7$, 
$\Delta E_{-}^{\ddagger} = 130.8$, 
$\Delta E_{+}^{\ddagger} = -135.2$, 
so that $\Delta E^{\ddagger}  = 266$, which is the same value obtained for $\Delta E^{\ominus}$ defined in equation~(\ref{kappa_eq_vanthoff}).
	One should note that, since $\Delta S^{\ddagger} = \beta^{*} \Delta E^{\ddagger}$, equation~(\ref{kappa_eq}) would retrieve the usual van't Hoff's expression, equation~(\ref{kappa_eq_vanthoff}), only if the free-energy profile $\beta^* \Delta F(E)$ present symmetric wells, that is, if $|\Delta E_{+}^{\ddagger}| = |\Delta E_{-}^{\ddagger}|$ and $\Delta \bar{\gamma} = 0$, which is not the case.
	Importantly, the equal height criteria based on the probability distribution $p(E)$~\cite{janke1998npB} is obtained when $\beta=\beta^*$, however, 
for non-symmetric wells, equation~(\ref{kappa_eq}) yields $k_{\text{eq}}(\beta^*)
= \Delta E_{-}^{\ddagger} / |\Delta E_{+}^{\ddagger}| \neq 1$,
so that $\kappa_-$ is slightly different from $\kappa_+$ at $\beta^*$.
	Alternatively, one may consider the equal area criteria
(see, e.g., reference~\cite{danielsson2015pnas})
where the equality $\kappa_-=\kappa_+$ (i.e.,~$k_{\text{eq}}=1$)
occur at a temperature $T_m=1/\beta_m$, with $\beta_m = \beta^* + \delta$ and $\delta
\approx (\Delta E^{\ddagger})^{-1} \ln \left( |\Delta E_{+}^{\ddagger}| /\Delta E_{-}^{\ddagger} \right)$.



	It is worth mentioning that, when $\delta \ll \beta^{*}$, 
 the ratios $\Delta S^{\ddagger}/\Delta E^{\ddagger}$ and $\Delta S^{\ominus}/\Delta E^{\ominus}$ will give the same inverse transition temperature $T^{*}$ evaluated from their microcanonical counterparts, i.e., $T^{*}=1/\beta^*$ computed from $\Delta S^{\dagger}$ and $\Delta E^{\dagger}$  as in equation~(\ref{transition_temperature_dagger}). 
	Even so, one should note that the estimates of the latent heat obtained from the fit of equations~(\ref{kappa_eq_vanthoff}) and~(\ref{kappa_eq}) to the numerical data 
present slightly smaller values, i.e.,~$\Delta E^{\ddagger} = \Delta E^{\ominus} = 266$, than the microcanonical latent heat, $\Delta E^{\dagger}=288$, which was evaluated directly from $S(E)$, as illustrated in figure~\ref{microcanonical_thermostatistics}(b).
	Table~\ref{parameters} shows that one have 
$\Delta E^{\ddagger} \approx 0.9 \, \Delta E^{\dagger}$
 also for other microcanonical entropies that were obtained with different parameters
and which display different free-energy barriers $\beta^* \Delta F^{\dagger}$.
	Thus, I note that such deviation is not particular of the parameters used
to obtain figure~\ref{microcanonical_thermostatistics}.
	In any case, one should recognize that it is a fair agreement considering 
all the approximations involved in the
derivation of equation~(\ref{kappa_eq}), and also that the
mean first-passage times were evaluated between the same energies $E_{-}$ and $E_{+}$
for all temperatures even though the maxima of $p(E)$ changed with $T$.

	Figure~\ref{fig_rate_constants}(a) indicates that 
the numerical results for the forward and the reverse rate constants obtained from the stochastic simulations are well fitted by expressions~(\ref{kappa_minus}) and~(\ref{kappa_plus}).
	Importantly, by leaving just $\gamma^{*}$ as a free parameter, that is, by considering the above values of $\Delta E_{-}^{\ddagger}$, $\Delta E_{+}^{\ddagger}$, with $\gamma_-$ and $\gamma_+$ consistent with the value $\Delta \bar{\gamma} $ given by equation~(\ref{kappa_eq}), and the free-energy barrier $\beta^* \Delta F^{\dagger}$ obtained from the microcanonical analysis, one have that $\gamma^{*} \approx 2 \times 10^{-4}$, which is consistent with the value one finds from the fit of the peak of the free-energy profile $\beta^{*} \Delta F(E)$ to the Gaussian function defined by equation~(\ref{betadeltaF_ast}) (data not shown).
	That value of $\gamma^{*}$ ensures a quantitative agreement even for the pre-factors
$A_-$ and $A_+$, which can be evaluated independently through equation~(\ref{prefactors}).

	In the following I validate my approach by applying the expressions derived in this section for the rate constants to describe the experimental data on the kinetics of first-order phase transitions of two molecular systems.

\subsection{Protein folding kinetics and protein stability}
\label{prot_fold_stab}

	As a first example, I consider the temperature-dependent kinetics related to protein folding-unfolding transitions~\cite{fershtbook}, where one can use several experimental techniques~\cite{danielsson2015pnas,gruebele2012pnas,schuler2013currop} to probe the refolding ($f$) and the unfolding ($u$) rate constants.
	Usually, the rate constants are determined in terms of Arrhenius-like expressions with the effective free-energies, also known as activation energies, defined as $\Delta G_{(u)f}^{\,}(T) = - k_BT \ln [\kappa_{(u)f}^{\,}(T)/\kappa_0^{\,}]$, so that~\cite{gruebele2012pnas}
\begin{equation}
\hspace{-2.5cm}
\kappa_{(u)f}^{\,}(T) \approx
\kappa_0^{\,} \,
\exp 
\left\{
-
\left[
 \Delta H_{(u)f}^{\,} - T \Delta S_{(u)f}^{\,}
+ \Delta C_{p}^{(u)f} \left( \left( T- T_m \right) -  T \ln \left( 
\frac{T}{T_{m}} 
\right) \right)
\right] \bigg/ k_B T \right\}~,
\label{pseudorates}
\end{equation}
where $T_m$ is a reference temperature, e.g.,
the midpoint of thermal denaturation~\cite{jackson1991biochem},
$\Delta H_{(u)f}^{\,}$ and $\Delta S_{(u)f}^{\,}$ are reference enthalpies and entropies at $T_m$, respectively, and $\Delta C_{p}^{(u)f}$ are estimates for the changes in heat capacity at constant pressure;
$\kappa_0^{\,}$ is an effective kinetic constant that sets the units of the rate constants and, in this case, is related to the viscosity of the solvent~\cite{gruebele2012pnas}.
	Alternatively, by assuming that the rate constants can be expressed by equations~(\ref{kappa_minus}) and~(\ref{kappa_plus}), one finds that the refolding (i.e.,~forward) and unfolding (i.e.,~reverse) rate constants can be written as
\begin{equation}
\hspace{-1.0cm}
\kappa_{(u)f}^{\,}(T)
\approx 
A_{0}^{\,} \, \exp \left\{ -
\left[
 \Delta E_{(u)f}^{\ddagger}  \left(1 - \frac{T}{T_m} \right)
+ \frac{\bar{\gamma}_{(u)f}^{\,} }{2 k_B T} \left( 1 - \frac{T}{T_m}  \right)^2 \,
\right] \bigg/ k_B T \right\}~~,
\label{kappa_fu_T}
\end{equation}
where the parameters $\Delta E_{(u)f}^{\ddagger}$
and $\bar{\gamma}_{(u)f}^{\,}$ are related, respectively, to the energy differences and curvatures that are obtained from the microcanonical free-energy profiles (see section~\ref{rate_const_sec}). 

\begin{figure*}[!t]
\centering
\includegraphics[width=0.96\textwidth]{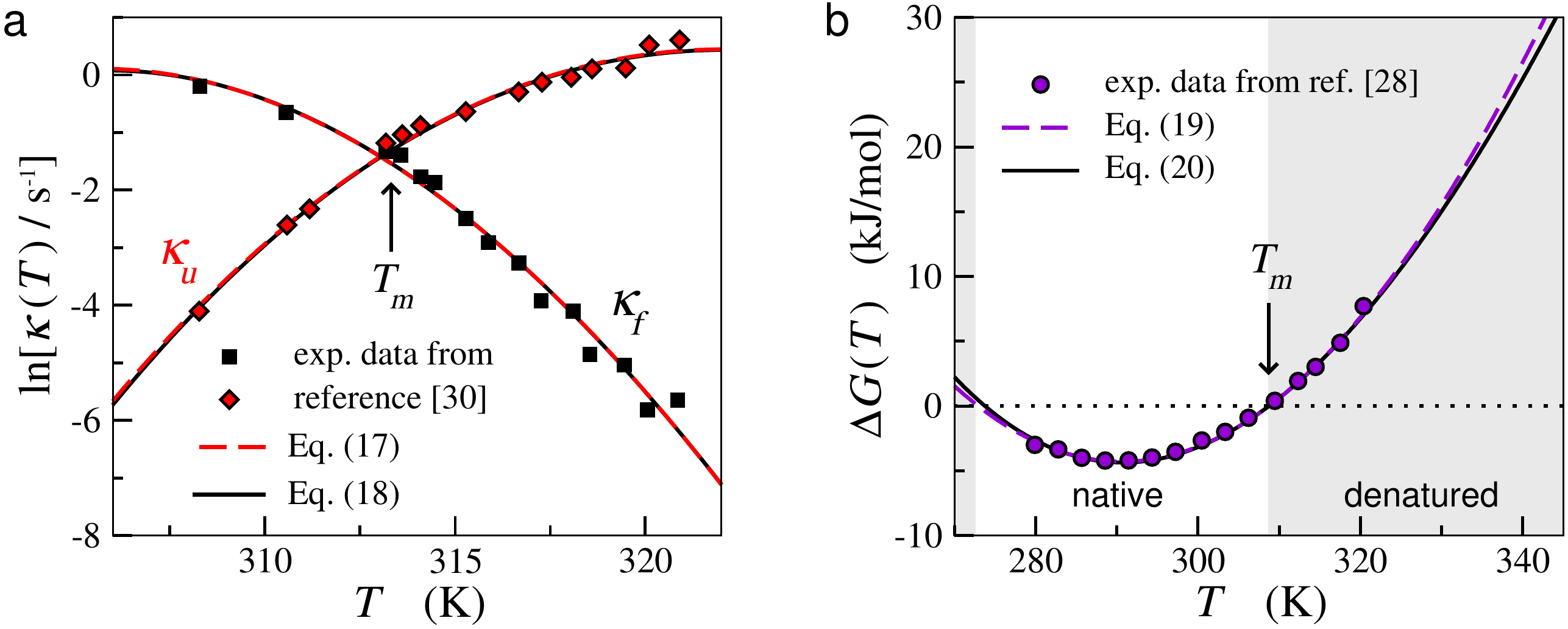}
\caption{ {\bf Refolding and unfolding rates and the protein folding stability.} 
(a) Filled symbols correspond to the experimental rate constants
extracted from reference~\cite{gruebele2012pnas}
for 
FRET-PGK protein.
	Dashed (red) lines correspond to equation~(\ref{pseudorates}) with
$T_m=313.05\,$K (i.e.,~$T_m=39.9^{\text{o}}$C) and $\kappa_0=0.24\,$s$^{-1}$
for both refolding ($\kappa_f^{\,}$) and unfolding ($\kappa_u^{\,}$)
rate constants;
for the refolding rate constant one have that
$\Delta H_f^{\,} = -333\,$kJ/mol,
$\Delta S_f^{\,} = -1.06\,$kJ/(mol.K),
and $\Delta C_p^f = -45\,$kJ/(mol.K);
while for the unfolding rate the values are
$\Delta H_u^{\,} = 345\,$kJ/mol,
$\Delta S_u^{\,} = 1.10\,$kJ/(mol.K),
$\Delta C_p^u = -37.8\,$kJ/(mol.K).
	Straight (black) lines correspond to 
equation~(\ref{kappa_fu_T}) with $A_0 = \kappa_0$;
for $\kappa_f^{\,}$ one have that
$\Delta E_{f}^{\ddagger} = -334\,$kJ/mol
and $\bar{\gamma}_f^{\,} = 37042\,$(kJ/mol)$^2$;
while for $\kappa_u^{\,}$ the values are
$\Delta E_{u}^{\ddagger} = 344.6\,$kJ/mol
and $\bar{\gamma}_u^{\,} = 31047\,$(kJ/mol)$^2$.
(b) Circles correspond to the experimental data for the free-energy $\Delta G(T)$ extracted from reference~\cite{danielsson2015pnas}
for SOD(I35A) enzyme in PBS buffer; 
the dashed (purple) line corresponds to 
equation~(\ref{deltaGoliveberg})
with $T_{m}=308.75\,$K (i.e.,~$T_m=35.6^{\text{o}}$C),
$\Delta H=-144.8\,$kJ/mol, $\Delta S=-0.47\,$kJ/(mol.K), and $\Delta C_p=-7.76\,$kJ/(mol.K);
and the continuous (black) line corresponds to equation~(\ref{deltaGrizzi})
with $\Delta E^{\ddagger} =-145.1\,$kJ/mol and
 $\Delta \bar{\gamma}=5827.3\,$(kJ/mol)$^2$.
Grey areas indicate regions that denatured states becomes 
favourable, including those related to cold denaturation~\cite{cooper2010jphyschemlett}
at low temperatures.
}
\label{fig_folding_stability}
\end{figure*}

	As indicated by the results presented in figure~\ref{fig_folding_stability}(a), the
expressions~(\ref{pseudorates}) and~(\ref{kappa_fu_T}) can describe equally well the temperature dependence of the rate constants obtained from experiments
on the folding-unfolding transition of the FRET-PGK protein~\cite{gruebele2012pnas}.
	By considering the values of the parameters obtained from the data displayed in figure~\ref{fig_folding_stability}(a), one can readily identify that the first two terms 
in the exponential of equation~(\ref{pseudorates}), which are related to the enthalpic and entropic contributions, should correspond to the first term in the exponential of equation~(\ref{kappa_fu_T}), so that $\Delta H_{(u)f}^{\,} \approx \Delta E_{(u)f}^{\ddagger}$ and $\Delta S_{(u)f} \approx \Delta S_{(u)f}^{\ddagger} = \Delta E_{(u)f}^{\ddagger}/T_m$.
	Importantly, despite the clear difference between the last terms in the exponentials of equations~(\ref{pseudorates}) and~(\ref{kappa_fu_T}), one can verify from the data presented in figure~\ref{fig_folding_stability}(a) that the ratio between those terms is almost constant over the range of temperatures considered (data not shown).
	Indeed, by assuming that $T$ is close to the transition temperature, i.e.,~$T\approx T_m$, one can use the approximation $\ln(1+x) \approx x - x^2/2$ with $x \approx (T-T_m)/T_m$, 
so that a comparison between equations~(\ref{pseudorates}) and~(\ref{kappa_fu_T}) 
indicates that $\Delta C_p^{(u)f} \approx - \bar{\gamma}_{(u)f}/k_B T_m^2$.


	It is worth mentioning that both forward and reverse rate constants given by equations~(\ref{pseudorates}) and~(\ref{kappa_fu_T}) should be interpreted as pseudo-equilibrium constants~\cite{tanford1968}, and are, in fact, related to relaxation kinetics of reversible processes close to the folding-unfolding transition~\cite{zwanzig1997pnas} (see~\ref{relaxation_rate_constant} for details).~The
	 actual equilibrium constant can be estimated experimentally as $\kappa_{\text{eq}}=[N]/[D]$ from measurements of the concentrations [N] and [D] of proteins (at equilibrium) in their native and denatured states, respectively (see, e.g.,~references~\cite{gruebele2012pnas,schuler2013currop,danielsson2015pnas}).
	Usually, it is assumed that the equilibrium constant can be written as
an Arrhenius-like expression, i.e.,~$\kappa_{\text{eq}}=\kappa_f/\kappa_u = e^{- \Delta G(T)/k_BT}$, with the corresponding effective free-energy given by~\cite{danielsson2015pnas,baldwin1986,oliveberg1995pnas}
\begin{equation}
\Delta G(T) 
= \Delta H - T \Delta S
+ \Delta C_p \left[ \left( T- T_m \right) -  T \ln \left( \frac{T}{T_{m}} \right) \right]~,
\label{deltaGoliveberg}
\end{equation}
where $\Delta H = \Delta H_{f}^{\,} - \Delta H_{u}^{\,}$,
$\Delta S= \Delta S_{f}^{\,} - \Delta S_{u}^{\,}$,
and $\Delta C_{p} = \Delta C_{p}^{f} - \Delta C_{p}^{u} $.
	Alternatively, the equilibrium constant $\kappa_{\text{eq}}$ can be evaluated from
the rate constants $\kappa_f$ and $\kappa_u$ defined by equation~(\ref{kappa_fu_T}), 
so that the protein folding stability is determined from an effective free-energy  
that is given by
\begin{equation}
\Delta G(T) = 
\Delta E^{\ddagger} \left(1 - \frac{T}{T_m} \right)
+ \frac{ \Delta \bar{\gamma}}{ 2 k_B T} \left(1 - \frac{T}{T_m} \right)^2 
~~,
\label{deltaGrizzi}
\end{equation}
where $\Delta E^{\ddagger} = \Delta E_{f}^{\ddagger} - \Delta E_{u}^{\ddagger}$
and 
$\Delta \bar{\gamma} = \bar{\gamma}_{f} - \bar{\gamma}_{u}$.
	I note that the derivation of the above equation assumes the equal area criteria, yet, one can consider equation~(\ref{kappa_eq}) to obtain a similar free-energy but assuming the equal height criteria (see the discussion in section~\ref{rate_const_sec}).

	Clearly, the results displayed in 
figure~\ref{fig_folding_stability}(b) indicate that
the temperature dependence of the free-energy $\Delta G(T)$ obtained from experiments on
SOD enzyme~\cite{danielsson2015pnas}
can  be well described by both equations~(\ref{deltaGoliveberg}) and~(\ref{deltaGrizzi}).
	Similarly to what is observed for the rate constants, 
one can identify that the first term in equation~(\ref{deltaGrizzi})
should correspond to the first two terms in equation~(\ref{deltaGoliveberg}), which are related to the enthalpic and entropic contributions,  so that $\Delta H \approx \Delta E^{\ddagger}$
and $\Delta S \approx \Delta S^{\ddagger}  = \Delta E^{\ddagger}/T^{*}$.
	And, again, one can verify that the ratio between the last terms in equations~(\ref{deltaGoliveberg}) and~(\ref{deltaGrizzi}) is almost constant for the data presented in 
figure~\ref{fig_folding_stability}(b), and, by comparing the approximated expressions,
one finds that $\Delta C_p \approx - \Delta \bar{\gamma}/k_B T_m^2$.
	In any case, one should note that the derivations of equations~(\ref{pseudorates}) and~(\ref{deltaGoliveberg}) 
consider that $\Delta C_p$ and $\Delta C_p^{(u)f}$ are independent of the temperature~\cite{oliveberg1995pnas}, and assume some approximations which are specific for a particular protein model~\cite{baldwin1986,jackson1991biochem}.
	Since the expressions~(\ref{kappa_fu_T}) and~(\ref{deltaGrizzi}) do not rely neither on that assumption or on any particular model, they might be useful to provide an interpretation to inconsistent results related to the protein folding-unfolding transition~\cite{cooper2010jphyschemlett}.

	In the context of computer simulations of protein folding phenomena, it is worth mentioning that, instead of the so-called {\it rugged free-energy landscapes}~\cite{jankebook}, the free-energy profiles $\beta^* \Delta F(E)$ that were obtained from replica exchange Monte Carlo simulations of heteropolymers display rather smooth curves~\cite{frigori2013jcp}, i.e.,~similar to what is presented in figure~\ref{microcanonical_thermostatistics}(b).
	One should also note that, although the microcanonical entropies $S(E)$ present a convex intruder region for systems which present first-order phase transitions (see, e.g., references~\cite{barre2001prl,frigori2010eurphysJB,frigori2010jphysconfser}), they are somewhat different from the {\it rugged funnel-like} picture commonly used to describe the folding-unfolding transitions, specially if one recall that, at finite temperatures, the native state of a protein does not necessarily corresponds to its ``ground-state''.

	Next I consider a problem that shares some features with the folding-unfolding transitions~\cite{fershtbook}, which is the nucleation phenomenon.






\subsection{Ice nucleation in supercooled water droplets}
\label{icenucleation}


	As discussed in references~\cite{murray2010physchemchemphys,murray2012chemsocrev},
experiments that measure nucleation rates of ice in supercooled water droplets 
at conditions similar to atmospheric cloud formation are very important to climate science.
	The state-of-the-art technique involve an ensemble 
with $M$, that is, thousands of micrometre-sized liquid droplets in contact with a 
cooling stage that works as a thermal reservoir.
	Hence, by assuming that, at a time $t^{\prime}$, $n_L(t^{\prime})$ droplets are in the liquid state at a temperature $T^{\prime}=T(t^{\prime})$, one can relate the number of droplets $\Delta n_F$ that should freeze after a interval of time $\Delta t = t^{\prime \prime} - t^{\prime}$ to the nucleation rate coefficient $J(T^{\prime})$ through an analytical expression that is given by~\cite{murray2010physchemchemphys,koop2013physchem,atkinson2016jphyschemA}
\begin{equation}
\Delta n_{F}(T^{\prime}) = n_F(t^{\prime \prime}) - n_F(t^{\prime}) = n_L(t^{\prime}) ( 1 -  e^{- J(T^{\prime}) V \Delta t })~~,
\label{deltanF}
\end{equation}
where $V$ is the 
volume attributed to a single liquid droplet.~In 
	practice, the experiments are done at a constant cooling rate, e.g., $r=-1\,$K/min, and 
one measures the fraction of frozen droplets $f_{F}(t)=n_F(t)/M$ at consecutive (discrete) times, i.e., $t^{\prime \prime} = t^{\prime} + \Delta t$.
	Since one can associate an average temperature $T^{\prime}$ to a given time interval $[t^{\prime},t^{\prime} + \Delta t[$, equation~(\ref{deltanF}) can be inverted to provide an experimental estimate for the nucleation rate coefficient, that is~\cite{koop2013physchem},
\begin{equation}
J(T^{\prime})  = - \frac{1}{V \Delta t} \ln \left[ \frac{1 -f_F(t^{\prime \prime})}{ 1 - f_F(t^{\prime}) } \right] ~~.
\label{exp_rate_coef}
\end{equation}
	
	Importantly, one should note that the stochastic description used to obtain 
equations~(\ref{deltanF}) and~(\ref{exp_rate_coef}) assumes that the whole droplet freezes
as a consequence of a single microscopic nucleation event~\cite{murray2012chemsocrev,koop2013physchem,atkinson2016jphyschemA}.
	This means that the nucleation rate $j(T)$, which is usually defined 
by an Arrhenius-like expression just like equation~(\ref{CNT_nucleation_rate}), 
can be directly related to the nucleation rate coefficient as
 $j(T) = J(T) V$~\cite{koop2013physchem}.
	Also, it turns out that the rate theory developed in section~\ref{rate_const_sec} is very convenient to describe those ice nucleation experiments, since the expressions for the rate constants derived in that section correspond to the phase transformation kinetics of the whole system, even though ``whole'' in this case (just as in computational simulations) means a very small portion of a micrometre-sized droplet.


	Because the nucleation experiments are performed with a varying temperature, one must
consider that the number of frozen droplets $n_F(t)$ at a time $t$ might not correspond to its expected equilibrium value. 
	Indeed, the time evolution of $n_F(t)$ should be governed by a relaxation
kinetics according to equation~(\ref{deltan_minus}), with an effective relaxation rate given by $J_{\text{obs}}$,
 just as discussed in~\ref{relaxation_rate_constant}.
	Hence, the number of droplets $\Delta n_F$ that will freeze after a interval of time $\Delta t$ at a temperature $T$ will be approximately given by equation~(\ref{deltanminus_approx}), which is identical to equation~(\ref{deltanF}), from
where one can identify that $J(T)V \approx J_-(T)V =  \kappa_{-}(T)$, with the
forward rate constant $\kappa_{-}$ given by equation~(\ref{kappa_minus}).
	Thus, by considering equation~(\ref{kobs_approx}), one can write the nucleation rate coefficient as a function of the temperature as
\begin{equation}
J(T) 
\approx J_{0} \,
\exp \left\{ 
- \left(\frac{1}{T} - \frac{1}{T^{*}} \right) \left[
\frac{\Delta E_{+}^{\ddagger}}{k_B}
+ \frac{\bar{\gamma}_{+}}{2 k_B^2} \left(\frac{1}{T} - \frac{1}{T^{*}} \right)
\right] \right\}
\label{nucleation_rate_jT}
\end{equation}
where
\begin{equation}
J_{0} 
\approx
\frac{D_{\varepsilon}}{ V  | \Delta E_{+}^{\ddagger} | } \sqrt{ \frac{\gamma^{*}}{2\pi} }
\, e^{- \beta^{*} \Delta F^{\dagger}}~~, 
\label{prefactors_LN}
\end{equation}
with the parameters defined as in section~\ref{rate_const_sec}.

\begin{figure*}[!t]
\centering
\includegraphics[width=0.98\textwidth]{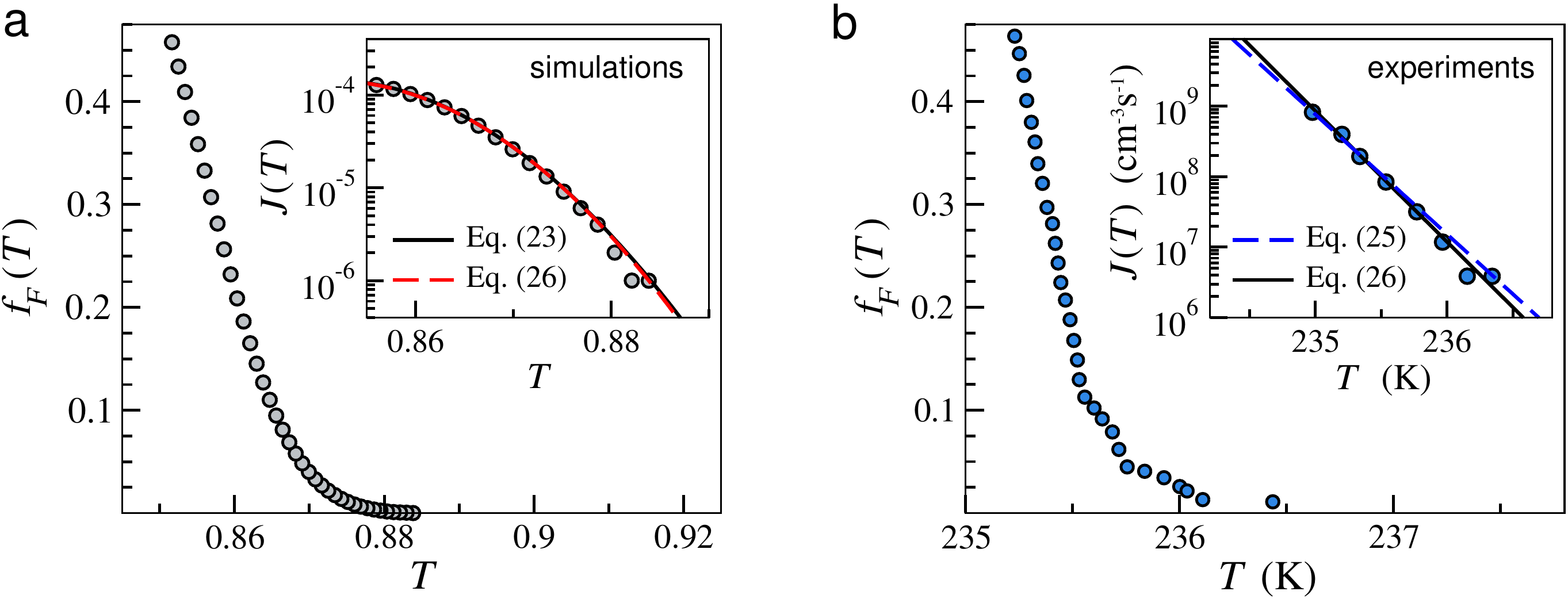}
\caption{
\footnotesize {\bf Ice nucleation rates in supercooled water droplets.} 
(a) Circles correspond to the fraction of frozen droplets $f_{F}^{\,}(T)$ as a
 function of the temperature $T$
obtained numerically with aid of equation~(\ref{deltan_minus})
by considering a relaxation rate given by $J_{\text{obs}}=(\kappa_- + \kappa_+)/V$,
with the rate constants $\kappa_-$ and $\kappa_+$ computed with
the parameters obtained from figure~\ref{fig_rate_constants}(a)
(it is also assumed that $M=10^4$, $V = 1$, $\Delta t=100$, and $k_B=1$).
Inset shows the nucleation rate coefficient:
circles correspond to $J(T)$ obtained from $f_{F}^{\,}(T)$ through
 equation~(\ref{exp_rate_coef}),
while the straight (black) and dashed (red) lines are
given by 
equations~(\ref{nucleation_rate_jT})
 and~(\ref{nucleation_rate_jT_approx})
with the same parameters used to generate the data in the main panel, i.e.,
$A_{+}=2.54\times 10^{-5}$, $\Delta E_{+}^{\ddagger}=-135.2$, and $\bar{\gamma}_{+}=5262.6$.
(b) Circles correspond to experimental data
on ice nucleation in micrometre-sized water droplets 
extracted from reference~\cite{atkinson2016jphyschemA}.
The main panel shows the fraction of frozen droplets $f_{F}^{\,}(T)$
while the inset include the nucleation rate coefficient
$J(T)$ evaluated from $f_{F}^{\,}(T)$ through equation~(\ref{exp_rate_coef}).
In the inset the straight (black) line corresponds to the fit to 
equation~(\ref{nucleation_rate_jT_approx}) by assuming
 $T^{*}=273.15\,$K and $\Delta E_{+}^{\ddagger}/k_B(T^{*})^2=-3.9126\,$K$^{-1}$, 
which yields
$\bar{\gamma}_{+} / (2 [k_B(T^{*})^2] ^2) \approx -0.01\,$K$^{-2}$ and $J_0 \approx 10^{-59}\,$cm$^{-3}$.s$^{-1}$,
%
while the dashed (blue) line corresponds to equation~(\ref{jT_phenomen})
with $a=-3.9126\,$K$^{-1}$ and $b=939.916\,$ (see reference~\cite{atkinson2016jphyschemA} for details). 
}
\label{fig_nucleation_rate}
\end{figure*}

	Figure~\ref{fig_nucleation_rate}(a) illustrates what one usually observes from ice nucleation experiments (see, e.g., figure~\ref{fig_nucleation_rate}(b)) and include numerical results that were produced in order to validate equation~(\ref{nucleation_rate_jT}).
	The main panel shows the fraction of replicas of the system that
are in the low energy phase (e.g.,~frozen droplets)
obtained with aid of equation~(\ref{deltan_minus}) by considering 
$J_{\text{obs}} = ( \kappa_{-} + \kappa_{+})/V$, with the rate constants given by the 
data displayed in figure~\ref{fig_rate_constants}(a) and 
arbitrary values of $V$ and $\Delta t$.
	As shown in the inset of figure~\ref{fig_nucleation_rate}(a), the nucleation rate $J(T)$ given by equation~(\ref{nucleation_rate_jT}) with the 
same parameters of the forward rate $\kappa_{-}$ displayed
in figure~\ref{fig_rate_constants}(a) 
is very similar to the numerical estimates evaluated from $f_F(T)$ through
equation~(\ref{exp_rate_coef}).
	Such result corroborates the idea discussed in~\ref{relaxation_rate_constant}
 that, indeed, 
the nucleation rate coefficient can be well approximated by the forward rate constant, that is,
$J(T) \approx J_{\text{obs}}(T)  \approx J_{-}(T)$.
	Interestingly, that result also justifies the use
of the forward rate, i.e., $J_{-} = \kappa_{-}/V$, as an estimate
for the nucleation rate used in numerical simulations~\cite{wedeking2006jcp}.


	In the context of atmospheric ice formation, one
often resort to empirical approaches~\cite{murray2010physchemchemphys}.~The 
	simplest phenomenological expression which is 
used by experimentalists to describe the nucleation rate coefficient 
is given by~\cite{koop2013physchem,atkinson2016jphyschemA}
\begin{equation}
\ln J(T) = a T + b ~~,
\label{jT_phenomen}
\end{equation}
with $a$ and $b$ defined as empirical parameters.~As 
	shown in figure~\ref{fig_nucleation_rate}(b), 
which includes the experimental data on homogeneous nucleation of ice in supercooled water droplets extracted from reference~\cite{atkinson2016jphyschemA}, equation~(\ref{jT_phenomen}) with $a=-3.9126\,$K$^{-1}$ and $b=939.916$ describes the data for $J(T)$ well (see the inset of that figure).
	In order to provide some reasoning behind that phenomenological expression,
one can assume that the temperature of the thermal reservoir is not too far 
from the transition temperature, i.e.,~$T \approx T^{*}$, 
so that the logarithm of equation~(\ref{nucleation_rate_jT}) 
can be approximately rewritten as
\begin{equation}
\ln J(T) 
\approx \ln J_{0}
+
\left(T - T^{*} \right) 
\left[
\frac{\Delta E_{+}^{\ddagger}}{k_B(T^{*})^2}
- \frac{\bar{\gamma}_{+}}{2 [k_B(T^{*})^2] ^2} \left( T - T^{*} \right) 
\right] ~~,
\label{nucleation_rate_jT_approx}
\end{equation}
where the parameters are the same as in equation~(\ref{nucleation_rate_jT}).
	As shown in the inset of the figure~\ref{fig_nucleation_rate}(b), 
equation~(\ref{nucleation_rate_jT_approx}) can be used to fit the experimental
equally well as equation~(\ref{jT_phenomen}).
	And, if one assumes that $\Delta E_{+}^{\ddagger} / k_B(T^{*})^2$
is given by the value of $a$ obtained from the empirical expression,
the value of the parameter $\bar{\gamma}_{+}$ determined by fitting the 
experimental data to equation~(\ref{nucleation_rate_jT_approx}) is very small 
(as shown in the inset figure).

	Unfortunately, the data available from the most of the experimental studies on homogeneous nucleation of ice are given only for a narrow range of temperatures just like in the figure~\ref{fig_nucleation_rate}(b), thus one cannot conclude (based solely on the fit) 
that the obtained parameters are reliable or not.
	In any case, here I use the numerical results evaluated from $J_{\text{obs}}=(\kappa_- + \kappa_+)/V$ in order to validate the approximated expression~(\ref{nucleation_rate_jT_approx}).
	Remarkably, as shown in the inset of the figure~\ref{fig_nucleation_rate}(a), 
equation~(\ref{nucleation_rate_jT_approx}) can be used to describe
the numerical data for $J(T)$ just as well as equation~(\ref{nucleation_rate_jT}), 
and, more importantly, with the same values for the parameters
$J_{0}$, $T^{*}$, $\Delta E_{+}$, and $\bar{\gamma}_{+}$.








	The main advantage of evaluate the nucleation rate from microcanonical free-energy profiles is that it does not require one to define an equimolecular dividing surface that separates the molecules that are in new phase from the molecules that are still in the old phase~\cite{dimobook,schmelzerbook}.
	As mentioned in the introduction, the microcanonical thermostatistics analysis can be considered a shape-free method, so it should not present any difficulties related to geometric features of the system that are shared by most of nucleation theories which are based on the classical nucleation theory.
	In the context of atmospheric ice nucleation~\cite{murray2010physchemchemphys}, in particular, equations~(\ref{nucleation_rate_jT}) and~(\ref{nucleation_rate_jT_approx}) should provide experimentalists an alternative way to analyse their data without having to resort to interfacial energies which are characterized by an arbitrary power-law dependence on $T$.





\section{Concluding remarks}


	Although inferences about the kinetics of first-order phase transitions based on microcanonical free-energy profiles
were suggested before in references~\cite{janke2017natcomm,frigori2013jcp}, microcanonical thermostatistics 
have been used mainly to describe the equilibrium properties of molecular systems (see, e.g., references~\cite{schnabel2011pre,junghans2006prl,moddel2010physchemchemphys,bereau2010jacs,church2012jcp,
alves2015cpc}).~In 
	this paper I have extended the use of microcanonical thermostatistics in order to develop a rate theory which provides simple temperature-dependent expressions for all rate constants, i.e.,~the forward ($\kappa_-$) and the reverse ($\kappa_+$) rate constants given by equations~(\ref{kappa_minus}) and~(\ref{kappa_plus}), respectively, 
and the equilibrium constant ($\kappa_{\text{eq}}$), which is given by equation~(\ref{kappa_eq}).
	Those expressions were validated through numerical results obtained from stochastic simulations, and I showed that $\kappa_-$ and $\kappa_+$ can display non-Arrhenius behaviours
(see figure~\ref{fig_rate_constants}(a)), just as it is observed experimentally for the kinetics data on protein folding-unfolding transitions presented in figure~\ref{fig_folding_stability}(a).


	Since the rate constants were derived from the mean first-passage times and those, in turn, were estimated  from the microcanonical free-energy profiles $\beta^{*}\Delta F(E)$, their analytical expressions should be useful in providing
the kinetics of finite molecular systems 
directly from their entropies $S(E)$, or inverse of the microcanonical temperatures $b(E)$, which are 
quantities that can be determined from several computational algorithms~\cite{berg2003cpc,pmc1996brazj,wanglandau2001prl,straub2011jcp,rizzi2011jcp}.~Importantly, 
	one should note that, although the microcanonical free-energy profiles 
used here were evaluated as a function of the energy 
$E$~(see, e.g., references~\cite{rizzi2011jcp,alves2015cpc}), 
it should be straightforward to generalize the expressions for the rates in terms of enthalpic-dependent entropies~\cite{straub2014prl,keyes2015jchemtheorcomput}.

	In addition, it is worth mentioning that in order to implement the stochastic simulations (see~\ref{stochasticsim}) as well as to derive the analytical expressions for the mean-first passage times (see~\ref{MFPTapp}), I have assumed that the stochastic processes are defined by one-dimensional Markov chains with an energy-independent diffusion coefficient $D_{\varepsilon}$.
	In order to extend the approach to stochastic processes that are obtained from more sophisticated computational simulations where the energy $E$ is a variable projected from a high-dimensional space, one might have to consider an energy-dependent diffusion coefficient $D(E)$ (see references~\cite{nadler2007pre,katzgraber2006jstatmech}). 
	Nevertheless, if the diffusion coefficient $D(E)$ present a
deep valley close to the transition energy $E^{*}$ (see, e.g.,~\cite{trebst2004pre}),
$D_\varepsilon$ might be effectively replaced by $D(E^*)$.
	Alternatively, one can pursuit an approach that involve a temperature-dependent diffusion coefficient $D(T)~$ (see, e.g.,~reference~\cite{bauer2010jstatmech}), but that is 
beyond the scope of the present work.

 

	Finally, one should note that,
because the microcanonical thermostatistics analysis is a shape-free 
method, the temperature-dependent expression for the nucleation rate coefficient $J(T)$, i.e.,~equation~(\ref{nucleation_rate_jT}) derived from the free-energy profiles, can be used to describe both homogeneous and heterogeneous nucleation processes.
	Also, 
further applications 
of the nucleation rate coefficients presented in section~\ref{icenucleation} might include experiments
that involve temperature-dependent kinetics
such as the time-temperature-transformation diagrams of
glass-based materials, e.g.,~glass-ceramics
and metallic glasses~\cite{zanotto2017intj},
as well as the aggregation kinetics of
biomolecules~\cite{knowles2014natrev,vekilov2016,rizzi2015jphyschemB,lair2020jphysconfser}.







\appendix






\section{~Microcanonical model for phase transitions}
\label{micromodel}

	Simplified models have been largely used in the study of phase transitions~\cite{thirring1970,janke2008prl,fiore2013jcp,matty2017physicaA}.
	Here I introduce an effective model to obtain $S(E)$  by assuming that its microcanonical temperature is given by
\begin{equation}
T(E) = -a_0 E + a_1 e^{b_1E} - a_2 e^{-b_2E} + a_3 ~~.
\label{tempE}
\end{equation}
	This function is used to evaluate the inverse of the microcanonical temperature through
the relation $b(E)=1/T(E)$.
	Importantly, both first-order and continuous phase transitions can be
 modelled by equation~(\ref{tempE}).
	First-order phase transitions are characterized by S-shaped 
microcanonical 
temperatures~\cite{frigori2013jcp,schnabel2011pre,barre2001prl,rizzi2016prl}
which can be obtained when the positive function $b(E)$ present an inflexion point at a positive 
energy value $E^*$.
	This energy can be found through the condition $d^2T(E)/dE^2=0$, which 
leads to $E^{*} = - (b_1+b_2)^{-1} \ln \left[ a_1 b_1^2 /  a_2 b_2^2 \right]$
or, equivalently, to the condition $a_1 b_1^2 <  a_2 b_2^2$.
	The curve $b(E)$ presented in figure~\ref{microcanonical_thermostatistics}(a) 
was obtained with
 $a_0=0.0011$,
 $a_1=0.02$,
 $b_1=0.005$,
 $a_2=1.2$,
 $b_2=0.01$,
 and $a_3=1.18$.
	Note that by choosing $a_3=a_2-a_1$ one have that ``ground-state'' energy is reached at zero absolute temperature, that is, $T(0)=0$.
	For those parameters, the transition temperature is 
$T^{*}=T(E^{*})=0.8704$ with $\beta^{*}=b(E^{*})=1/T^{*}=1.1489$.
	Table~\ref{parameters} list the parameters that can be used to generate 
different microcanonical temperatures
that display S-shaped curves and which can be used to describe first-order phase
transitions. 
	In order to perform the stochastic simulations explained below,
I consider a energy discretization $E_m=E_0 + m \varepsilon$, with $E_0=0$, 
$\varepsilon=2$, and $m=0,1,\dots \,\,$.
	Hence, the microcanonical entropy is estimated from the values of
$b(E_m)$ using a piece-wise relation $S(E_m) = b(E_m) E_m - a(E_m)$, where the values 
of $a(E_m)$ are determined from $b(E_m)$ through the recurrence relations of the 
multicanonical algorithm~\cite{berg2003cpc,rizzi2011jcp,alves2015cpc}.

\begin{table}[!h]
\caption{\label{parameters}Parameters used to generate the different microcanonical temperatures $T(E)$ defined by equation~(\ref{tempE}); inverse of the microcanonical transition temperature, $\beta^{*}$;
free-energy barrier, $\beta^{*} \Delta F^{\dagger}$;
microcanonical latent heat, $\Delta E^{\dagger}$;
latent heat obtained from equation~(\ref{kappa_eq}), $\Delta E^{\ddagger}$;
}
\footnotesize\rm
\begin{tabular*}{\textwidth}{ccccc}
\br
Parameters  & $\beta^{*}$ &   $\beta^{*} \Delta F^{\dagger}$ & $\Delta E^{\dagger}$ 
& $\Delta E^{\ddagger}$ \\
\mr

 $a_0=0.0011$, $a_1=0.02$, $b_1=0.005$, $a_2=1.2$, $b_2=0.01$               &  $1.1489$            & $1.16$ & $288$ &  $266$ \\
 $a_0=0.0009$, $a_1=0.02$, $b_1=0.0048$, $a_2=1.12$, $b_2=0.012$              &  $1.1441$           &  $1.53$ & $320$ &  $286$  \\
 $a_0=0.001 $, $a_1=0.02$, $b_1=0.0049$, $a_2=1.2$, $b_2=0.012$              &  $1.0859$            &  $2.15$ & $350$  &  $316$ \\
 $a_0=0.001 $, $a_1=0.02$, $b_1=0.00495$, $a_2=1.2$, $b_2=0.013$               &  $1.0747$           &  $2.74$ & $370$ &  $338$   \\
 $a_0=0.0013$, $a_1=0.02$, $b_1=0.0053$, $a_2=1.27 $, $b_2=0.011$              &  $1.1180$           &  $3.30$ & $360$ & $330$  \\

\br
\end{tabular*}
\end{table}




\section{~Stochastic simulations}
\label{stochasticsim}

\begin{figure}[!t]
\centering
\includegraphics[width=0.98\textwidth]{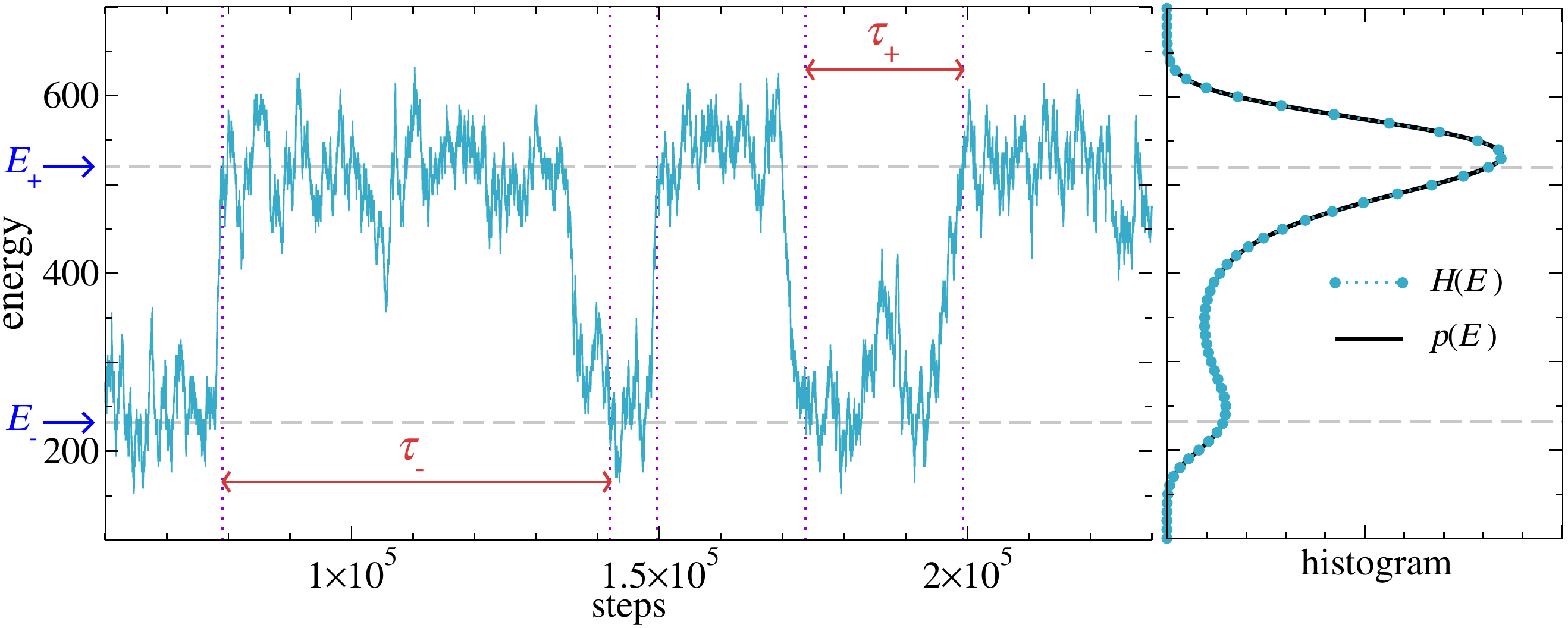}
\caption{
{\bf Markov chain and energy histogram obtained from stochastic simulations.}~Left panel include the 
time series produced by the stochastic protocol described in~\ref{stochasticsim}
for the microcanonical entropy $S(E)$ displayed in figure~\ref{microcanonical_thermostatistics}(b), 
for a temperature $T=0.875$, 
which is just above the transition temperature $T^{*}=0.8704$ 
(here it is assumed that $k_B=1$).
Horizontal dashed (grey) lines indicate the energies $E_{-}=232$ and $E_{+}=520$
which are used to evaluate the first-passage times, which are denoted
by $\tau_-$ and $\tau_+$ (see reference~\cite{trebst2004pre} for details).
Right panel shows the energy histogram $H(E)$ obtained from the time series with
$N_s = 3 \times 10^9\,$steps, while the straight (black) line correspond to the
canonical distribution $p(E)$ given by Eq.~(\ref{canonical_prob}).
}
\label{simulations_series}
\end{figure}

From the microcanonical entropy $S(E_m)$ at a discretized value of energy $E_m$, 
one can determine the density of states $\Omega(E_m)=e^{S(E_m)}$ and
evaluate the equilibrium canonical distribution as
$p(E_m) = [\mathcal{Z}(\beta)]^{-1} \Omega(E_m) e^{-\beta E_m}$.
	It is worth mentioning that special care is needed
in the evaluation of the partition 
function $\mathcal{Z}(\beta)$, since it requires
the large number summation technique that is described
in reference~\cite{berg2003cpc}.
	Also, in order to avoid numerical instabilities,
a threshold value of $p_{\text{min}} = 10^{-10}$ is used to
set the range where $p(E)>0$, that is defined between the energies $E_{i}$ (initial) and $E_{f}$ (final).	
	Those values of energy are determined from the probabilities $p(E_{i})$ and 
$p(E_{f})$ which are just above the threshold value $p_{\text{min}}$.
	The kinetics of the system is simulated from 
microscopic transitions using a simple stochastic 
approach, where the system with energy $E_m$ go to a 
energy $E_{m+1}$ with probability $T_{m,m+1}$
or to a energy $E_{m-1}$ with probability $T_{m,m-1}$. 
	The transition probabilities $T_{m,n}$ define a stochastic
matrix and are obtained from the equilibrium 
distribution $p_m = p(E_m)$.
	First, the transition probabilities are defined for $m=i$, that is,
$T_{i,i}=\Omega(E_{i})/[\Omega(E_{i}) + \Omega(E_{i+1})]$ and
$T_{i,i+1}=1- T_{i,i}$. 
	Then one should consider the detailed balance condition, so that $T_{i+1,i}=(p_{i}/p_{i+1})T_{i,i+1}$.
	After that, the rest of the transition probabilities in the stochastic matrix can be updated by considering the following steps:
(i) $T_{m+1,m} = (p_{m} - p_{m-1}T_{m-1,m} )/p_{m+1}$, which comes from
the equilibrium distribution;
(ii) $T_{m,m+1}=(p_{m+1}/p_{m})T_{m+1,m}$, which comes from the detailed balance condition;
then, steps (i) and (ii) are repeated from $m=i+1$ to $m=f-1$.
	Finally, when $m=f$ one have $T_{f+1,f}=0$ so that $T_{f,f}=1-T_{f,f-1}$.
	Each stochastic simulation at a given temperature corresponds to 
$N_{s}=3\times10^{9}$ steps starting at a random energy within the interval $[E_i,E_f]$.
	As shown in Fig.~\ref{simulations_series}, 
the Markov chain produced by such stochastic protocol lead to a energy histogram $H(E)$ 
that is equivalent to the canonical distribution $p(E)$ given by Eq.~(\ref{canonical_prob}).




\section{~Mean first-passage times}
\label{MFPTapp}

	As mentioned in section~\ref{rate_const_sec}, the mean first-passage times $\tau_{+}$ and $\tau_{-}$ are numerically evaluated from the stochastic simulations by considering the method of labelled walkers described in reference~\cite{trebst2004pre} (see figure~\ref{simulations_series} for details).
	In order to obtain their analytical expressions, I consider
the approach described in reference~\cite{nadler2007pre}, where
the mean-first passage time for the system 
to go from $E_{+}$ to $E_{-}$ is given by
\begin{equation}
\tau_{-} = \frac{1}{D} \int_{E_{-}}^{E_{+}} \frac{dE}{p(E)}
\int_{E}^{E_{+}} p(E') dE' ~~,
\end{equation}
where $D=\varepsilon^2/2\tau_\varepsilon$ is an energy-independent diffusion coefficient,
$\varepsilon$ is the bin size, and $\tau_\varepsilon$ determines the time scale involved in a microscopic energy transition.
	By following a similar approach discussed in reference~\cite{gardinerbook}, 
one can conveniently evaluate those integrals as
\begin{equation}
\tau_{-} \approx \frac{2 \tau_\varepsilon}{\varepsilon^2} 
\int_{-\infty}^{\infty} \frac{dE}{p^{*}(E)}
\int_{E^{*}}^{E_{+}} p_{+}(E') dE' \approx 
\sqrt{ \frac{8 \pi \tau_{\varepsilon}^2 (\Delta E_{+}^{\ddagger})^2}{\varepsilon^4\gamma^{*}}}
 \frac{\Gamma_{+}(\beta)}{\Gamma^{*}(\beta)}~~,
\label{tau_minus}
\end{equation}
where I assume that the error function can be approximated by erf($z$) $\approx z$, and $p^*(E)$ and $p_{+}(E)$ are given by equations~(\ref{prob_ast}) and~(\ref{prob_plus_minus}), respectively.
	Similarly, the mean first-passage time for the system to go from $E_{-}$ to $E_{+}$ is given by
\begin{equation}
\tau_{+} \approx 
\sqrt{ \frac{8 \pi \tau_{\varepsilon}^2 (\Delta E_{-}^{\ddagger})^2}{\varepsilon^4\gamma^{*}}}
 \frac{\Gamma_{-}(\beta)}{\Gamma^{*}(\beta)}~~,
\label{tau_plus}
\end{equation}
with $\Gamma^{*}(\beta)$ and $\Gamma_{-}(\beta)$ given by equations~(\ref{gamma_ast})
and~(\ref{gamma_plus_minus}), respectively.

\section{~Relaxation rate and the nucleation rate coefficient}
\label{relaxation_rate_constant}

	Consider an ensemble with $M$ replicas of the system in contact
with a thermal reservoir at a temperature $T$.
	Each replica has a fixed volume $V$ and the system can be either in a (thermodynamic) phase ``$+$'' or in a phase ``$-$'', each which might denote, respectively,
denatured and native states of a protein (if the system corresponds to an isolated protein in solution just as exemplified in section~\ref{prot_fold_stab}), or, liquid and frozen phases (if the system corresponds to a water droplet, as in the experiments discussed in section~\ref{icenucleation}).
	By assuming that the interconversion between the two thermodynamic phases can be described by a reversible reaction of the kind 
$X_{+} \rightleftharpoons X_{-}$, with the forward (``$+$'' $\rightarrow$ ``$-$'') and the reverse (``$-$'' $\rightarrow$ ``$+$'') 
rate constants given by~\cite{wedeking2006jcp} $J_{-} = \kappa_{-}/V$ and $J_{+} = \kappa_{+}/V$, respectively, one have that the fraction $f_{-}(t)$ of replicas in the phase
``$-$'' changes in time according to the following flux balance equation~\cite{zhou2010review,gardinerbook}
\begin{equation}
\frac{d f_{-}(t)}{dt} = - J_{+} V f_{-}(t) +  J_{-} V f_{+}(t) ~~.
\end{equation}
	If the ensemble have a fraction $f_{-}(t_0)$ of the replicas
in the phase ``$-$'' at a time $t_0$, after a interval of time $\Delta t$ this
fraction will be given by
\begin{equation}
f_{-}(t_0 + \Delta t) = 
  f_{-}(t_0) \, e^{ - J_{\text{obs}} V \Delta t}
+ f_{-}^{\text{eq}} \, (1 -  e^{ - J_{\text{obs}} V \Delta t} )
\label{solution_fminus}
\end{equation}
where 
\begin{equation}
J_{\text{obs}} = J_{-} + J_{+} ~~,
\label{jobs}
\end{equation}
and 
\begin{equation}
f_{-}^{\text{eq}} = \frac{J_-}{J_{\text{obs}} } = \frac{\kappa_{\text{eq}}}{ 1 + \kappa_{\text{eq}}}
\label{fminuseq}
\end{equation}
is the equilibrium fraction of replicas in the phase ``$-$'', with 
the equilibrium constant given by $\kappa_{\text{eq}} = J_{-}/J_{+}$.

	At equilibrium (and also for stationary states), one can verify through equation~(\ref{solution_fminus}) that if $f_{-}(t_0) \approx f_{-}^{\text{eq}}$ then
$f_{-}(t_0 +\Delta t)\approx f_{-}(t_0)$, which means that $df_{-}/dt \approx 0$.~However, 
	if $f_{-}(t_0 +\Delta t) \neq f_{-}^{\text{eq}}$ one can use equation~(\ref{solution_fminus}) to estimate the change in number of replicas in the phase ``$-$'' after a time $\Delta t$ as
\begin{equation}
\hspace{-1.0cm}
\Delta n_{-}(\Delta t) = 
M [  f_{-} (t_0 + \Delta t) - f_{-} (t_0) ]
=
[ n_{-}^{\text{eq}} -  n_{-}(t_0)] \, (1 -  e^{ - J_{\text{obs}} V \Delta t} ) ~~,
\end{equation}
where $n_{-}^{\text{eq}} = M f_{-}^{\text{eq}}$
and $n_{-}(t_0) = M f_{-}(t_0)$.
	Since, at any time, $n_{-} + n_{+} = M$, the above equation
can be rewritten as
\begin{equation}
\Delta n_{-}(\Delta t) =
n_{+}(t_0) \left[ 1 - \frac{n_{+}^{\text{eq}}}{n_{+}(t_0)} \right] \, (1 -  e^{ - J_{\text{obs}} V \Delta t} ) ~~,
\label{deltan_minus}
\end{equation}
where 
\begin{equation}
n_{+}^{\text{eq}} = M f_{+}^{\text{eq}} = \frac{M}{ 1 + \kappa_{\text{eq}}}
\label{npluseq}
\end{equation}
is the equilibrium number of replicas in the phase ``$+$''.

	Note that, if $n_{+}(t_0)$ is different from $n_{+}^{\text{eq}}$,
one can see from equation~(\ref{deltan_minus}) that
$\Delta n_{-}(\Delta)$ will be not zero and, in that case, the fraction of replicas in the phase
``$-$''  will relax to equilibrium at a rate $J_{\text{obs}}$
according to equation~(\ref{solution_fminus}).~The
	 relaxation rate $J_{\text{obs}}$, also known as the rate coefficient, is the main quantity measured in experiments on relaxation kinetics, e.g.,~ice nucleation~\cite{murray2010physchemchemphys} and protein folding~\cite{gruebele2012pnas} experiments.

	In practice, one might consider that the temperature of the thermal reservoir which the replicas are immersed is very low in comparison to the transition temperature (e.g.,~a liquid droplet in the supercooled region), so that
\begin{equation}
\ln  J_{\text{obs}} 
=
\ln \left( J_- + J_+ \right)
=
\ln \left[ J_- \left(  1  +  \kappa_{\text{eq}}^{-1} \right) \right] 
\approx
\ln J_-  
 ~~ , 
\label{Jobs_full}
\end{equation}
where it is assumed that 
$\ln(1+\kappa_{\text{eq}}^{-1})\approx 0$, since
the equilibrium constant $\kappa_{\text{eq}} \gg 1$ for temperatures below the transition temperature  $T^{*}$, i.e.,~$\beta>\beta^{*}$.
	In particular, for the case of microcanonical thermostatistics discussed in section~\ref{rate_const_sec}, one can obtain the relaxation rate constant $J_{ \text{obs} }\approx J_{-}$ by considering the forward rate constant, $\kappa_- = J_- V $, which is given by equation~(\ref{kappa_minus}), so that
\begin{equation}
\ln J_{\text{obs}}   
\approx
\ln J_{-} 
\approx \ln J_{0} 
- (\beta - \beta^{*}) \left[
\Delta E_{+}^{\ddagger}
+ \frac{\bar{\gamma}_{+}}{2} (\beta - \beta^{*})
\right]
\label{kobs_approx}
\end{equation}
with $J_{0}$ defined as in equation~(\ref{prefactors_LN}).
	In addition, at the supercooled region, i.e.,~$T<T^*$, one have from equation~(\ref{npluseq}) 
that the (expected) equilibrium number of replicas  in the phase with higher energy will be small, i.e.,~$n_{+}^{eq} \ll M$, 
hence one can approximate equation~(\ref{deltan_minus}) to
\begin{equation}
\Delta n_{-}(\Delta t) \approx
n_{+}(t_0) \, (1 -  e^{ - J_{-} V \Delta t} ) ~~,
\label{deltanminus_approx}
\end{equation}
which is an expression that is identical to equation~(\ref{deltanF}).
	As discussed in section~\ref{icenucleation}, equation~(\ref{deltanF}) can be used to determine the nucleation rate coefficient $J(T)$ from experiments on ice nucleation in supercooled water droplets~\cite{murray2010physchemchemphys}.
	By comparing equations~(\ref{deltanminus_approx}) and~(\ref{deltanF}) one can readily identify that the nucleation rate can be approximated by the forward rate constant, that is, $j(T) = J(T) V \approx J_{-}(T)V = \kappa_-(T)$.









\noindent
\section*{Acknowledgements}


	I am especially grateful to Professors Dimo Kashchiev, Nelson Alves, and Erich Meyer, 
for the inestimable knowledge they shared with me. 
	I also thank the financial support of the Brazilian 
agencies CNPq (Grants N\textsuperscript{o} 306302/2018-7 and N\textsuperscript{o} 426570/2018-9) and FAPEMIG (Process APQ-02783-18), although no funding from FAPEMIG
was released until the submission of the present work.











\section*{{\bf References}}
\addcontentsline{toc}{section}{{\bf References}}



\begin{thebibliography}{10}
\expandafter\ifx\csname url\endcsname\relax
  \def\url#1{{\tt #1}}\fi
\expandafter\ifx\csname urlprefix\endcsname\relax\def\urlprefix{URL }\fi
\providecommand{\eprint}[2][]{\url{#2}}

\bibitem{knowles2014natrev}
Knowles T~P~J, Vendruscolo M and Dobson C~M 2014 {\em Nat. Rev. Mol. Cell
  Biol.\/} {\bf 15} 384

\bibitem{vekilov2016}
Vekilov P~G 2016 {\em Progress in Crystal Growth and Characterization of
  Materials\/} {\bf 62} 136

\bibitem{zhou2010review}
Zhou H~X 2010 {\em Q. Rev. Biophys.\/} {\bf 43} 219

\bibitem{eyring1935jcp}
Eyring H 1935 {\em J. Chem. Phys.\/} {\bf 3} 107

\bibitem{kramers1940phys}
Kramers H~A 1940 {\em Physica VII\/} {\bf 4} 284

\bibitem{cooper2010jphyschemlett}
Cooper A 2010 {\em J. Phys. Chem. Lett.,\/} {\bf 1} 3298

\bibitem{dimobook}
Kashchiev D 2003 {\em Nucleation: Basic Theory with Applications\/}
  (Butterworth-Heinemann)

\bibitem{schmelzerbook}
Schmelzer J~W~P 2005 {\em Nucleation Theory and Applications\/} (Wiley-VCH)

\bibitem{cabriolu2012jcp}
Cabriolu R, Kashchiev D and Auer S 2012 {\em J. Chem. Phys.\/} {\bf 137} 204903

\bibitem{bingham2013jcp}
Bingham R~J, Rizzi L~G, Cabriolu R and Auer S 2013 {\em J Chem Phys.\/} {\bf
  139} 241101

\bibitem{grossbook}
Gross D~H~E 2001 {\em Microcanonical Thermodynamics\/} (World Scientific)

\bibitem{janke2017natcomm}
Zierenberg J, Schierz P and Janke W 2017 {\em Nat. Commun.\/} {\bf 8} 14546

\bibitem{frigori2013jcp}
Frigori R~B, Rizzi L~G and Alves N~A 2013 {\em J. Chem. Phys.\/} {\bf 138}
  015102

\bibitem{truhlar2001pnas}
Truhlar D~G and Kohen A 2001 {\em Proc. Natl. Acad. Sci. USA\/} {\bf 98} 848

\bibitem{meyer1986jcrysgrow}
Meyer E 1986 {\em J. Cryst. Growth\/} {\bf 74} 425

\bibitem{schnabel2011pre}
Schnabel S, Seaton D~T, Landau D~P and Bachmann M 2011 {\em Phys. Rev. E\/}
  {\bf 84} 011127

\bibitem{frenkelbook}
Frenkel D and Smit B 2002 {\em Understanding Molecular Simulation\/} (Academic
  Press)

\bibitem{berg2003cpc}
Berg B~A 2003 {\em Comput. Phys. Commun.\/} {\bf 153} 397

\bibitem{pmc1996brazj}
de~Oliveira P~M~C, Penna T~J and Herrmann H~J 1996 {\em Braz. J. Phys.\/} {\bf
  26} 677

\bibitem{wanglandau2001prl}
Wang F and Landau D~P 2001 {\em Phys. Rev. Lett.\/} {\bf 86} 2050

\bibitem{straub2011jcp}
Kim J, Keyes T and Straub J~E 2011 {\em J. Chem. Phys.\/} {\bf 135} 061103

\bibitem{rizzi2011jcp}
Rizzi L~G and Alves N~A 2011 {\em J. Chem. Phys.\/} {\bf 135} 141101

\bibitem{janke1998npB}
Janke W 1998 {\em Nuclear Physics B\/} {\bf 63} 631


\bibitem{leekosterlitz1990prl}
Lee J and  and Kosterlitz J~M 1990 {\em Phys. Rev. Lett.} {\bf 65} 137


\bibitem{crooks2000pre}
Crooks G~E 2000 {\em Phys. Rev. E\/} {\bf 61} 2361

\bibitem{krivov2013pre}
Krivov S~V 2013 {\em Phys. Rev. E\/} {\bf 88} 062131

\bibitem{hanggi1999pre}
Reimann P, Schmid G~J and H\"anggi P 1999 {\em Phys. Rev. E\/} {\bf 60} R1

\bibitem{trebst2004pre}
Trebst S, Huse D~A and Troyer M 2004 {\em Phys. Rev. E 70\/} {\bf 70} 046701

\bibitem{danielsson2015pnas}
Danielsson J, Mu X, Lang L, Wang H, Binolfi A, Theillet F~X, Bekei B, Logan
  D~T, Selenko P, Wennerstr\"om H and Oliveberg M 2015 {\em Proc. Natl. Acad.
  Sci. USA\/} {\bf 112} 12402

\bibitem{fershtbook}
Fersht A 1999 {\em Structure and Mechanism in Protein Science\/} (W. H. Freeman
  and Company)

\bibitem{gruebele2012pnas}
Guo M, Xu Y and Gruebele M 2012 {\em Proc. Natl. Acad. Sci. USA\/} {\bf 109}
  17863

\bibitem{schuler2013currop}
Schuler B and Hofmann H 2013 {\em Curr. Opin. Struct. Biol.\/} {\bf 23} 36

\bibitem{jackson1991biochem}
Jackson S~E and Fersht A~R 1991 {\em Biochemistry\/} {\bf 30} 10428

\bibitem{tanford1968}
Tanford C 1968 {\em Adv. Protein Chem.\/} {\bf 23} 121

\bibitem{zwanzig1997pnas}
Zwanzig R 1997 {\em Proc. Natl. Acad. Sci. USA\/} {\bf 94} 148

\bibitem{baldwin1986}
Baldwin R~L 1986 {\em Proc. Natl. Acad. Sci. USA\/} {\bf 83} 8069

\bibitem{oliveberg1995pnas}
Oliveberg M, Tan Y~J and Fersht A~R 1995 {\em Proc. Natl. Acad. Sci. USA\/}
  {\bf 92} 8926

\bibitem{jankebook}
Janke W 2008 {\em Rugged Free Energy Landscapes. Lect. Notes Phys. 736\/}
  (Springer)

\bibitem{barre2001prl}
Barr\'e J, Mukamel D and Ruffo S 2001 {\em Phys. Rev. Lett.\/} {\bf 87} 030601

\bibitem{frigori2010eurphysJB}
Frigori R~B, Rizzi L~G and Alves N~A 2010 {\em Eur. Phys. J. B: Cond. Matt.
  Phys.\/} {\bf 75} 311

\bibitem{frigori2010jphysconfser}
Frigori R~B, Rizzi L~G and Alves N~A 2010 {\em J. Phys.: Conf. Ser.\/} {\bf
  246} 012018

\bibitem{murray2010physchemchemphys}
Murray B~J, Broadley S~L, Wilson T~W, Bull S~J, Wills R~H, Christenson H~K and
  Murray E~J 2010 {\em Phys. Chem. Chem. Phys.\/} {\bf 12} 10380

\bibitem{murray2012chemsocrev}
Murray B~J, O'Sullivan D, Atkinson J~D and Webb M~E 2012 {\em Chem. Soc.
  Rev.\/} {\bf 41} 6519

\bibitem{koop2013physchem}
Riechers B, Wittbracht F, H\"utten A and Koop T 2013 {\em Phys. Chem. Chem.
  Phys.\/} {\bf 15} 5873

\bibitem{atkinson2016jphyschemA}
Atkinson J~D, Murray B~J and O'Sullivan D 2016 {\em J. Phys. Chem. A\/} {\bf
  120} 6513

\bibitem{wedeking2006jcp}
Wedekind J, Reguera D and Strey R 2006 {\em J. Chem. Phys.\/} {\bf 125} 214505

\bibitem{junghans2006prl}
Junghans C, Bachmann M and Janke W 2006 {\em Phys. Rev. Lett.\/} {\bf 97}
  218103

\bibitem{moddel2010physchemchemphys}
M\"oddel M, Janke W and Bachmann M 2010 {\em Phys. Chem. Chem. Phys.\/} {\bf
  12} 11548

\bibitem{bereau2010jacs}
Bereau T, Bachmann M and Deserno M 2010 {\em J. Am. Chem. Soc.\/} {\bf 132}
  13129

\bibitem{church2012jcp}
Church M~S, Ferry C~E and van Giessen A~E 2012 {\em J. Chem. Phys.\/} {\bf 136}
  245102

\bibitem{alves2015cpc}
Alves N~A, Morero L~D and Rizzi L~G 2015 {\em Comput. Phys. Commun.\/} {\bf
  191} 125

\bibitem{straub2014prl}
Cho W~J, Kim J, Lee J, Keyes T, Straub J~E and Kim K~S 2014 {\em Phys. Rev.
  Lett.\/} {\bf 112} 157802

\bibitem{keyes2015jchemtheorcomput}
Malolepsza E and Keyes T 2015 {\em J. Chem. Theory Comput.\/} {\bf 11} 5613

\bibitem{nadler2007pre}
Nadler W and Hansmann U~H~E 2007 {\em Phys. Rev. E\/} {\bf 75} 026109

\bibitem{katzgraber2006jstatmech}
Katzgraber H~G, Trebst S, Huse D~A and Troyer M 2006 {\em J. Stat. Mech.\/}
  P03018

\bibitem{bauer2010jstatmech}
Bauer B, Gull E, S T, Troyer M and Huse D~A 2010 {\em J. Stat. Mech.\/}  P01020

\bibitem{zanotto2017intj}
Zanotto E~D, Tsuchida J~E, Schneider J~F and Eckert H 2015 {\em Int. Mater.
  Rev.\/} {\bf 60} 376

\bibitem{rizzi2015jphyschemB}
Rizzi L~G and Auer S 2015 {\em J. Phys. Chem. B\/} {\bf 119} 14631

\bibitem{lair2020jphysconfser}
Trugilho L~F and Rizzi L~G 2020 {\em J. Phys.: Conf. Ser.\/} {\bf 1483} 012011


\bibitem{thirring1970}
Thirring W 1970 {\em Z. Physik\/} {\bf 235} 339

\bibitem{janke2008prl}
Bittner E, Nu{\ss}baumer A and Janke W 2008 {\em Phys. Rev. Lett.\/} {\bf 101}
  130603

\bibitem{fiore2013jcp}
Fiore C~E and da~Luz M~G~E 2013 {\em J. Chem. Phys.\/} {\bf 138}

\bibitem{matty2017physicaA}
Matty M, Lancaster L, Griffin W and Swendsen R~H 2017 {\em Physica A\/} {\bf
  467} 474

\bibitem{rizzi2016prl}
Rizzi L~G and Alves N~A 2016 {\em Phys. Rev. Lett.\/} {\bf 117} 239601

\bibitem{gardinerbook}
Gardiner C 2004 {\em Handbook of Stochastic Methods\/} 3rd ed (Springer)

\end{thebibliography}

\providecommand{\newblock}{}

\end{document}